# Job insecurity, equilibrium determinacy and E-stability in a New Keynesian model with asymmetric information. Theory and simulation analysis

Luca Vota[1*] and Luisa Errichiello[2]

## Abstract

Departing from the dominant approach focused on individual and meso-level determinants, this manuscript develops a macroeconomic formalization of job insecurity within a New Keynesian framework in which the IS–NKPC–Taylor rule block is augmented with labor-market frictions. The model features partially informed private agents who receive a noisy signal about economic fundamentals from a fully informed public sector. When monetary policy satisfies the Taylor principle, the equilibrium is unique and determinate. However, the release of news about current or future fundamentals can still generate a form of "Paradox of Transparency" through general-equilibrium interactions between aggregate demand and monetary policy. When the Taylor principle is violated, belief-driven equilibria may emerge. Validation exercises based on the Simulated Method of Moments confirm the empirical plausibility of the model's key implications.

## 1. Introduction

Precarious work is a particularly widespread phenomenon globally, both in advanced economies and in emerging and developing countries. According to a survey conducted by the Ipsos institute on 12,000 adult workers residing in 27 countries and presented at the World Economic Forum in 2020 (Ipsos 2020), 54% of the sample declared themselves concerned about the possibility of losing their job in the 12 months following the interview (of these, 17% defined themselves as "very concerned", while 37% said they were "fairly concerned"). The highest values were recorded in Russia (75%), Spain (73%), and Malaysia (71%), while the lowest ones were reported in Germany (26%), Sweden (30%), the Netherlands, and the United States (36%). Although they indicate strong variability between countries, these data show that job insecurity, defined as the subjectively perceived probability of experiencing an interruption in one's career (Shoss, 2017), is a highly pervasive issue even in countries with a strong industrial vocation that offer their workers generous welfare measures (as in the case of Germany and Sweden).

Starting from the 1990s, precisely because of the structural changes that have affected economies globally and that have made traditional relationships between workers and businesses less structured and more flexible (Davis, 2013), social scientists have shown a growing interest in job insecurity. Factors such as globalization and automation have made a decisive contribution to reducing the costs and times of production activities but, at the same time, have increased workers' perception of insecurity (Mughan et al., 2003; Scheve & Slaughter, 2006; Couple, 2019; Nam, 2019; Raeder et al., 2019; Cao & Song, 2025). Contrary to what one might imagine, technological progress does not constitute a threat solely to the career continuity of low-skilled workers, but also to that of individuals with high professional qualifications (Colvin, 2015). More recent evidence produced from qualitative and quantitative studies has highlighted that the introduction of Artificial Intelligence (AI) has significantly influenced perceived job insecurity, without relevant differences between employees in

---
[1*] Corresponding author. Institute for Studies on the Mediterranean of the National Research Council (CNR-ISMed), Naples, Italy. luca.vota@cnr.it
[2] Institute for Studies on the Mediterranean of the National Research Council (CNR-ISMed), Naples, Italy. luisa.errichiello@cnr.it

managerial roles and those in non-managerial roles (Koo et al., 2021). Others, however, have discovered that the negative impact of AI on job insecurity is mediated by vocational learning ability, i.e., the ability to autonomously acquire and apply new skills to one's own work environment (Liu & Zhan, 2020). In fact, the need to interact with AI to perform their duties encourages workers to constantly seek to improve their skills, with important implications for their insecurity, creativity, well-being, and psychological health (Wu et al., 2024). The fear that AI will rapidly make their skills obsolete, leading to the replacement of human operators with virtual ones, is another important source of stress for workers (Sharif et al., 2025).

Existing research to date has favored a micro and meso approach focused on individual, job-related, and organizational antecedents of job insecurity, paying little attention to systemic, macroeconomic ones. In particular, studies have used survey data relating to variables such as contract type, responsibilities associated with one's position, hourly wage, health conditions, number of children, age, years of experience, and so on (De Witte, 2005; Clark & Postel-Vinay, 2009; Böckerman et al., 2011).

Such analyses have often focused on specific sectors and regions to account for the heterogeneity present in different economic sectors and in the economic conditions of various countries, proving useful in evaluating the predictors of job insecurity in different contexts (Lee et al., 2018; Chirumbolo et al., 2020; Martínez et al., 2020; De Cuyper et al., 2021; Ibanescu et al., 2023; Muñoz Medina et al., 2023; Darvishmotevali, 2025).

However, a fundamental limitation of the individual-based approach is its inability to account for macroeconomic variables and (consequently) the influence that the alternation of the various phases of the business cycle exerts on job insecurity. Indeed, while purely individual variables undoubtedly have a significant impact on job insecurity, it is equally plausible that a non-negligible component of insecurity is predicted by macroeconomic variables such as actual GDP, output gap, inflation, unemployment rate, investment, household consumption and saving, and so on. As demonstrated by the literature on Dynamic Stochastic General Equilibrium (DSGE) models, these variables move along one or multiple equilibrium paths in response to exogenous shocks that are more or less persistent and potentially capable of triggering phases of expansion and recession (Dave & Sorge, 2025). Governments and central banks react to cyclical fluctuations by using the economic policy tools at their disposal (taxation, public spending, social transfers, interest rates, and money supply) to stabilize price and national income growth, preserve employment, and contain the harmful effects of adverse shocks. This complex macroeconomic dynamic inevitably reflects on the expectations of households and firms about the fundamentals and future perspectives of being unemployed.

The lack of studies focused on the macroeconomic dimension does not allow for discerning the short-run and long-run effects of exogenous shocks on job insecurity, nor for defining public policies to combat job insecurity that adequately account for the effects of the business cycle. In fact, to effectively design and implement their policies for containing job insecurity, policymakers need analytical tools that consider the gradual processes of divergence and convergence of the actual unemployment rate in relation to its steady-state equilibrium (which, in turn, are strictly dependent on cyclical fluctuations of the economy).

The paper's original purpose is to formalize job insecurity under the alternative information hierarchies (full information and asymmetric information) and policy regimes (satisfaction versus violation of the Taylor principle) and thus theoretically and empirically assess the impacts of the diverse model's shocks on this variable.

To accomplish these tasks, the paper proposes a small-scale, rational expectations New Keynesian model populated by public agents (central bank and government) who are fully informed and private agents (firms and households) who directly observe only a part of the state variables and receive an informative signal comprising the news on the current and future values of the fundamentals and an uninformative noise capturing communication ambiguity and forecasting errors.

The model comprises the usual dynamic IS curve (IS)-New Keynesian Phillips curve (NKPC)-contemporaneous Taylor rule block and a set of equations for the law of motion of the unemployment rate, job-finding intensity, and job-separation intensity that are consistent with the modern modelling of the labor market (Petrongolo & Pissarides, 2001; Shimer, 2005; Fujita & Ramey, 2009; Gertler & Trigari, 2009).

The informative signal is modeled as an exogenous shifter of the dynamic IS curve: the news shocks hit aggregate demand and then propagate to the entire economy, also triggering the response of the central bank.

The baseline version of the model is extended to account for cohort heterogeneity with the further aim of providing fresh insights on the macroeconomic determinants of the mature workers' job insecurity (a subject matter to which scholars are paying increasing attention).

The research study reveals that, under both the full and asymmetric information regimes, the satisfaction of the Taylor principle (namely, the policy conduct according to which the central bank responds aggressively to an increase in prices by raising its policy rate more than proportionally to the inflation rate) remains the necessary condition to ensure equilibrium determinacy, uniqueness, and E-stability under decreasing-gain learning. When the Taylor principle is violated, the asymmetric information version of the theoretical model may admit belief-driven (sunspot) equilibria anchored to innovations in the informative signal; nevertheless, such equilibria are not learnable and therefore are not selected by adaptive learning dynamics.

In all the informative and monetary policy schemes, the short-run persistent news shocks substantially contribute to the definition of the equilibrium paths of the endogenous variables. Remarkably, it is found that the net impact of a positive news shock about the current and future state of the economy on the expected job-separation intensity is critically dependent on the tension between its demand effect and the policy reaction of the central bank. In this sense, it is possible to say that the model is subject to the "Paradox of Transparency" phenomenon (in the sense of Morris & Shin): the transmission of otherwise unobservable information by public agents to private ones is not necessarily beneficial for the perceived job security.

The theoretical results have been validated on Italian quarterly time series ranging between 2004Q1 and 2025Q2. The reason for choosing Italy is the availability of aggregate and age-specific data, which makes it possible to empirically assess the theoretical results of both the aggregate version of the model and its extension incorporating cohort heterogeneity. The data related to mature workers are collected from the Survey on Health, Ageing and Retirement in Europe (SHARE), which gathers individual-level statistics from 28 different countries starting from 2004. Italy represents the most appropriate case study among the countries surveyed by SHARE because, according to the data released by the United Nations in 2022, it is the member country of the Organization for Economic Cooperation and Development (OECD) whose population displays the highest median age after Japan (United Nations, 2022).

The rest of the paper is organized as follows: Section 2 proposes a literature review on job insecurity, with a clear indication of the gap filled by this research work; Section 3 presents the rational expectations New Keynesian model with asymmetric information; Section 4 presents the full information solution; Section 5 characterizes the model's equilibrium under asymmetric information; Sections 6 and 7 evaluate the robustness of the rational expectations equilibria to the alternative hypothesis of adaptive learning; Section 8 extends the baseline version of the model to account for cohort heterogeneity; Sections 9 and 10 summarise the simulation strategy designed to qualitatively assess the model's theoretical predictions; Section 11 lists the time series used in the study; Section 12 contains the results of the validation; Section 13 discusses the obtained theoretical results; and finally, Section 14 draws the conclusions of the study and some policy recommendations on how to proficiently curb job insecurity. A detailed bibliography closes the paper.

## 2. Job insecurity research: moving toward a systemic approach

At its core, job insecurity is the perceived threat that one's job will not continue, a forward-looking construct widely examined across sociology, economics, and organizational psychology (Heaney,

Israel, & House, 1994; Shoss, 2017). By distinguishing these perceptions from realized labor-market events, the literature lays the groundwork for more fine-grained distinctions.

First, objective vs. subjective job (in)security differentiates realized or externally observed risk (e.g., contract type, displacement) from personal expectations and worries about job continuity (De Witte & Näswall, 2003; Erlinghagen, 2008). Second, quantitative vs. qualitative insecurity distinguishes perceived threats to the job itself (quantitative) from threats to valued features of the job (qualitative), such as career prospects, learning opportunities, and pay growth (Hellgren et al., 1999; Urbanaviciute et al., 2021). Third, cognitive vs. affective insecurity separates likelihood judgments ("How likely am I to lose my job?") from emotional reactions ("How worried am I?") (Borg & Elizur, 1992; Huang et al., 2012).

Measurement mirrors these distinctions. Objective indicators include contract status (temporary/permanent) and unemployment spells and displacement, and institutional features like employment protection. These are prevalent in economic analyses linking structural conditions to security outcomes, whereas macro-comparative work often aggregates objective proxies at the country level (Burgard et al., 2009; De Witte & Näswall, 2003; Green & Leeves, 2013). Subjective insecurity, widely used in organizational research and sociology, overwhelmingly uses perceptual instruments, and job (in)security is usually captured via multi-item survey scales, typically Likert-type items (e.g., "I might lose my job in the next year") (Ashford et al., 1989; Hellgren, 1999; Vander Elst et al., 2014), and many have undergone reliability and invariance testing across languages and countries (Lee et al., 2008; Görgens-Ekermans et al., 2024).

Within organizational psychology and sociology traditions, a large body of work documents antecedents and consequences of subjective job insecurity. Meta-analytic evidence shows that job insecurity is linked to poorer mental and physical health, reduced job satisfaction and commitment, higher turnover intentions, and, in many cases, diminished performance (Sverke et al., 2002; Sverke et al., 2019; Hur, 2022).

Antecedents operate at multiple levels. Across various literature reviews (Keim et al., 2014; Shoss, 2017; Lee, Huang & Ashford, 2018; Jian et al., 2021), they cluster into: institutional/macro factors (e.g., labor-market institutions, unemployment); organizational factors (e.g., restructuring/downsizing, union presence); job/position factors (e.g., temporary or contingent contracts, part-time status, lower-skill/blue-collar roles); individual demographics (e.g., tenure, minority status, education); individual experiential factors (e.g., perceived employability, training, prior unemployment); personality/core self-evaluations (e.g., locus of control, affectivity, self-esteem); and interpersonal/social factors (e.g., bullying, leadership, coworker/supervisor support). Focusing on Jiang et al. (2021), their multilevel meta-analysis adopts a Job Demands–Resources lens, recasting determinants as resources (personal and organizational) that reduce job insecurity and demands (personal and organizational) that increase it - showing that resources exert the stronger overall effects, underscoring the protective impact of employability, fair procedures, participation, and supportive leadership.

Cross-disciplinary studies also highlight that subjective perceptions are not epiphenomenal: they only partly overlap with objective conditions and can exert independent effects on well-being and behavior (De Witte & Näswall, 2003; Burgard et al., 2009). For example, temporary work does not uniformly depress attitudes, whereas subjective insecurity robustly predicts lower satisfaction and commitment across institutional settings (De Witte & Näswall, 2003). At the same time, research has begun to differentiate cognitive and affective facets, with the latter (worry) sometimes mediating or amplifying the effects of the former (likelihood judgments) on strain and performance (Jiang & Lavaysse, 2018; Huang et al., 2012). Relatedly, literature on employability suggests that perceived external options can buffer insecurity's effects - though not uniformly across contexts or populations (Fugate et al., 2004; de Cuyper et al., 2014; Yeves et al., 2019).

A second, largely sociological and economic stream scales up the analysis to the country level, examining whether macro-institutions and context shape average insecurity levels and their distribution. Drawing on extant research, Errichiello & Falavigna (2024) propose a framework grouping environmental factors shaping job insecurity into two broad groups: institutional categories and socio-economic structural factors. Institutions cover the following areas: (i) labour market; ii) social protection; iii) workers' power; iv) national culture; and, v) governance. Studies leverage correlational, multilevel models along with secondary data and cross-national surveys to relate insecurity to: active labour market policies (ALMPs) (Chung & Van Oorschot, 2011; Inanc & Kalleberg, 2022; Lübke & Erlinghagen, 2014); employment protection legislation (EPL) (Balz, 2017; Inanc & Kalleberg, 2022); social protection measures (Erlinghagen, 2008; Inanc & Kalleberg, 2022), workers' power (union density and collective bargaining) (Esser & Olsen, 2012; Inanc & Kalleberg, 2022), national cultural values (Erlinghagen, 2008; Moy et al., 2023; Errichiello & Falavigna, 2024), and quality of governance (e.g., Dixon et al., 2013; Errichiello & Falavigna, 2024) Regarding structural labour-market and socio-economic factors, empirical research correlates job insecurity to both current and long-term high unemployment (Chung & van Oorschot, 2011), GDP growth rate (Lübke & Erlinghagen, 2014) and country-level income inequality (Errichiello & Falavigna, 2025). This work typically compares countries or time periods around crises (e.g., 2008), reporting that macroeconomic slack, institutional protections, and cultural-institutional contexts correlate with perceived insecurity - yet with mixed and sometimes modest effect sizes across outcomes and subpopulations.

Despite their contributions, existing studies exhibit two limitations germane to the present paper's aims. First, most are correlational and rely on static or quasi-static designs (cross-sectional snapshots or pooled panels). This makes it difficult to model transmission mechanisms - for example, how shocks to aggregate demand, policy interventions (ALMPs, EPL reforms), or changes in governance quality propagate into workers' expectations about job continuity over time. Second, even when macro variables are included, these models rarely have explicit predictive structure: they do not incorporate expectation formation, feedback loops between unemployment/output and perceptions, or the role of policy rules, all central to New-Keynesian (NK) frameworks. Consequently, they are limited in forecasting subjective job insecurity under counterfactual monetary-fiscal or ALMP/EPL scenarios, or in evaluating the dynamic welfare and distributional consequences of policy regimes.

These gaps are consequential because subjective quantitative insecurity is precisely where macro shocks meet micro behavior. Recent reviews argue that insecurity is shaped by both demands (e.g., reorganization, role overload, macro slack) and resources (e.g., communication, employability, institutional buffers), with resources often exerting stronger protective effects (Jiang et al., 2021). Embedding this insight in a macro model requires a structural mapping from aggregate states (output gap, unemployment, policy rates, inflation, fiscal stance, etc.) to agents' expectations of job loss, allowing those expectations to influence consumption, labor supply, and wage setting - canonical NK channels. Doing so would extend macro-comparative work that links insecurity to structural socio-economic factors and institutions (Erlinghagen, 2008; Chung & van Oorschot, 2011; Lübke & Erlinghagen, 2014) by providing a dynamic, forward-looking, and policy-counterfactual apparatus that can forecast trajectories of subjective risk, not just correlate them with realized outcomes.

A large body of empirical literature shows that workers' expectations about job loss and labor-market risks are misaligned with the economic fundamentals, as they display systematic misperceptions, excessive volatility, and strong heterogeneity (Dominitz & Manski, 1997; Hendren, 2017; Mueller & Spinnewijn, 2023). Similarly, workers tend to overstate their subjectively perceived job-separation probabilities, especially in conditions of marked uncertainty and in response to noisy or ambiguous informative signals about the state of the economy (Roth & Wohlfart, 2020; Faberman et al., 2022). The analysis conducted in this manuscript is consistent with this empirical evidence and investigates its deeper determinants by establishing a formal theoretical basis for the macroeconomic dimension of job insecurity under diverse information hierarchies (including asymmetric information).

### 3. A New Keynesian model with asymmetric information and labor market

Consider an economy populated by private agents (households and firms) who optimize their respective expected profit and expected utility functions, a central bank that has a dual mandate (to stabilize the actual inflation rate and minimize the output gap), and a fiscal policy authority (government) that collects resources through taxation and re-employs them through government spending.

It is assumed that the central bank and the government are fully informed about the endogenous variables (actual output, output gap, actual inflation, government spending, public debt, nominal interest rate, real interest rate, unemployment rate, job-finding intensity, and job-separation intensity), exogenous variables (potential output, natural interest rate, steady-state job-finding intensity, and steady-state job-separation intensity), and exogenous shocks (shock to potential output, cost-push shock to inflation, news shock about the current state of the economy, news shock about the future values of the fundamentals, shock to government spending, demand shock, monetary shock, shock to job-finding intensity, and shock to job-separation intensity) affecting the economy. Private agents can directly observe only the current values of the nominal interest rate ($i_t$), actual inflation rate ($\pi_t$), real government debt stock ($d_t$), informative signal released by the public sector ($a_t$), and all the past values of the model's endogenous variables. By forming their expectations about future inflation, households and firms infer the market interest rate ($r_t$).

The informative signal $a_t$ is composed of three terms which are not separately observable to them: the information released by government and central bank about the current state of the economic fundamentals ($q_t$), the information on the future values of the fundamentals ($\varepsilon_t^b$), and the uninformative noise due to ambiguity and communication errors ($v_t$).

The full information set is defined as $\mathcal{F}^t = \{\mathcal{F}_t, \mathcal{F}_{t-1}, \dots\}$, namely, it includes the entire history of the observables contained in:

$$\mathcal{F}_t = \left\{ a_t, q_t, \varepsilon_t^b, v_t, \lambda_t, \xi_t, y_t, \hat{y}_t, \bar{y}_t, \bar{f}_t, \bar{s}_t, \omega_t, \varepsilon_t^{\hat{y}}, \eta_t, r_t, \bar{r}_t, \pi_t, u_t, f_t, s_t, \varepsilon_t^f, \varepsilon_t^s, \epsilon_t, \varsigma_t, i_t, \varepsilon_t^i, w_t, d_t, g_t, \vartheta_t \right\}$$

Let $\mathcal{H}^t = \{\mathcal{H}_t, \mathcal{H}_{t-1}, \mathcal{H}_{t-2}, \dots \mathcal{H}_{t-i}, \dots, \mathcal{H}_{t-I}\}$ be the private agents' information set, where $\mathcal{H}_{t-i} \subset \mathcal{F}_{t-i}$ with $i = 0, \dots, I$ is a vector whose unique non-null elements are given by the nominal interest rate ($i_t$), market interest rate ($r_{t-i}$), inflation rate ($\pi_{t-i}$), government debt ($d_{t-i}$), and informative signal ($a_{t-i}$).

Upon labeling with subscript "$t$" the variables observed by all the model's agents and with subscript "$t|t$" the variables that are not directly observed by the private sector, the informative signal can be formalized as in equation (1):

$$a_t = q_{t|t} + \varepsilon_{t|t}^b + v_{t|t} \tag{1}$$

where $v_{t|t} \sim iid(0, \sigma_v^2)$ is orthogonal to the fundamentals. Both $q_{t|t}$ and $\varepsilon_{t|t}^b$ follow stationary AR(1) processes:

$$q_{t|t} = \rho_q q_{t-1} + \lambda_{t|t} \tag{2}$$

and:

$$\varepsilon_{t|t}^b = \rho_b \varepsilon_{t-1}^b + \xi_{t|t} \tag{3}$$

where $|\rho_q| < 1$, $|\rho_b| < 1$, while $\lambda_{t|t} \sim iid(0, \sigma_\lambda^2)$ and $\xi_{t|t} \sim iid(0, \sigma_\xi^2)$ are, respectively, the informative shock to the current values of the fundamentals and the informative shock about the future state of the economy.

Actual output $y_t$ is given by the sum between output gap $\hat{y}_t$ and potential output $\bar{y}_t$:

$$y_{t|t} = \hat{y}_{t|t} + \bar{y}_{t|t} \tag{4}$$

where, of course, $\bar{y}_{t|t}$ follows a stationary AR(1) process (Smets & Wouters, 2007):

$$\bar{y}_{t|t} = \rho_{\bar{y}}\bar{y}_{t-1} + \omega_{t|t} \tag{5}$$

with $|\rho_{\bar{y}}| < 1$ and $\omega_{t|t} \sim iid(0, \sigma_\omega^2)$.

The aggregate demand of the private sector (households and firms) is expressed by the dynamic IS curve:

$$\hat{y}_{t|t} = E_t[\hat{y}_{t+1|t}|\mathcal{H}_t] - \frac{1}{\sigma}(\hat{\imath}_{t|t} - E_t[\hat{\pi}_{t+1|t}|\mathcal{H}_t] - \bar{r}_{t|t}) + \delta_a a_{t|t} + \delta_G \hat{g}_{t|t} + \varepsilon_{t|t}^{\hat{y}} \tag{6}$$

where $\hat{y}_{t|t}$ is the output gap, $\hat{\pi}_t$ is the deviation of the actual inflation rate from the constant inflation target set by the central bank ($\hat{\pi}_t = \pi_t - \bar{\pi}$), $\hat{\imath}_{t|t}$ is the deviation of the nominal interest rate from the inflation target ($\hat{\imath}_t = i_t - \bar{\pi}$), $\hat{g}_{t|t}$ is the deviation of government spending from its steady-state value ($\hat{g}_{t|t} = g_{t|t} - \bar{g}_{t|t}$), $\bar{r}_{t|t}$ is the natural interest rate, and $\varepsilon_{t|t}^{\hat{y}}$ is the demand shock admitting the following stationary AR(1) representation:

$$\varepsilon_{t|t}^{\hat{y}} = \rho_{\hat{y}}\varepsilon_{t-1}^{\hat{y}} + \eta_{t|t} \tag{7}$$

with $|\rho_{\hat{y}}| < 1$ and $\eta_{t|t} \sim iid(0, \sigma_\eta^2)$. Since they are both exogenous shifters of the dynamic IS curve, $\varepsilon_{t|t}^b$ is assumed to be orthogonal to $a_t$ (this avoids weak identification between demand and information).

The difference between the current nominal interest rate and the expected future inflation rate defines the real interest rate as in the standard Fisher equation:

$$\hat{r}_t = \hat{\imath}_t - E_t[\hat{\pi}_{t+1|t}|\mathcal{H}_t] \tag{8}$$

while the natural interest rate $\bar{r}_{t|t}$ is dictated by the Wicksell equation:

$$\bar{r}_{t|t} = \sigma(E_t[\bar{y}_{t+1|t}|\mathcal{H}_t] - \bar{y}_{t|t}) \tag{9}$$

Actual inflation obeys the New Keynesian Phillips curve:

$$\hat{\pi}_t = \beta E_t[\hat{\pi}_{t+1|t}|\mathcal{H}_t] + k\hat{y}_{t|t} + \epsilon_{t|t} \tag{10}$$

where $k$ is the slack parameter (slope), $\epsilon_{t|t}$ is the cost-push shock to inflation following a stationary AR(1) process (Christiano et al., 2005):

$$\epsilon_{t|t} = \rho_\epsilon \epsilon_{t-1} + \varsigma_{t|t} \tag{11}$$

with $|\rho_\epsilon| < 1$ and $\varsigma_{t|t} \sim iid(0, \sigma_\varsigma^2)$.

The central bank uses the nominal interest rate as its sole policy instrument and sets it on the basis of the standard contemporaneous Taylor rule:

$$\hat{\imath}_t = \alpha_\pi \hat{\pi}_t + \alpha_y \hat{y}_{t|t} + \varepsilon_{t|t}^i \tag{12}$$

where $\varepsilon_{t|t}^i$ is a stationary AR(1) monetary shock:

$$\varepsilon_{t|t}^i = \rho_i \varepsilon_{t-1}^i + w_{t|t} \tag{13}$$

with $|\rho_i| < 1$ and $w_{t|t} \sim iid(0, \sigma_w^2)$.

The unemployment rate updates between time $t-1$ and time $t$ as a result of the hiring and firing decisions of firms (transition intensities) according to the following law of motion:

$$u_{t|t} = (1 - f_{t|t})u_{t-1} + s_{t|t}(1 - u_{t-1}) \tag{14}$$

where job-finding intensity $f_{t|t}$ and job-separation intensity $s_{t|t}$, defined as:

$$f_{t|t} = 1 - e^{-\bar{f}_{t|t}} \tag{15}$$

and:

$$s_{t|t} = 1 - e^{-\bar{s}_{t|t}} \tag{16}$$

are determined at time $t$ and are applied to the transition between time $t-1$ and time $t$.

In accordance with Petrongolo & Pissarides (2001), Shimer (2005), Fujita & Ramey (2009), and Gertler & Trigari (2009), job-finding intensity $f_{t|t}$ and job-separation intensity $s_{t|t}$ follow stationary ARX(1) processes that additionally control for the output gap $\hat{y}_{t|t}$ and the deviation of the real interest rate $r_t$ from its natural level $\bar{r}_{t|t}$, namely:

$$f_{t|t} = \bar{f}_{t|t} + \rho_f(f_{t-1} - \bar{f}_{t-1}) + \phi_y \hat{y}_{t|t} - \phi_r(\hat{r}_t - \bar{r}_{t|t}) + \varepsilon^f_{t|t} \tag{17}$$

and:

$$s_{t|t} = \bar{s}_{t|t} + \rho_s(s_{t-1} - \bar{s}_{t-1}) - \psi_y \hat{y}_{t|t} + \psi_r(\hat{r}_t - \bar{r}_{t|t}) + \varepsilon^s_{t|t} \tag{18}$$

where $\bar{f}_{t|t}$ and $\bar{s}_{t|t}$ are the values of the long-run, steady-state job-finding and job-separation intensities compatible with potential output, $|\rho_f| < 1$, $|\rho_s| < 1$, while $\varepsilon^f_{t|t} \sim iid(0, \sigma^2_f)$ and $\varepsilon^s_{t|t} \sim iid(0, \sigma^2_s)$ are the white noise shocks to the transition intensities.

In line with the standard search-matching theory, in equation (17), job-finding intensity increases in the output gap and decreases in the real interest rate in deviation. Hence, the higher the aggregate demand, the higher the profit opportunities for firms, and the higher the vacancies. On the contrary, the higher the cost of financing investment and circulating capital, the higher the cost of posting a vacancy. These relationships are inverted when looking at the job-separation intensity in equation (18).

Real government debt ($\frac{D_t}{P_t} = d_t$) evolves over time according to the linearized dynamic constraint below:

$$\hat{d}_t = \hat{d}_{t-1} + \bar{d}\hat{r}_{t-1} \tag{19}$$

where $\hat{d}_t$ and $\hat{r}_t$ are the deviations of the real public debt and real interest rate from their respective steady-state values, while $\bar{d}$ is the steady-state level of $d_t$. The linearization of government debt in equation (19) avoids the need to explicitly control for GDP growth and preserves linearity in the model's solution.

Government spending (or, more precisely, the deviation of government spending from its steady-state value) follows a stationary AR(1) process (Leeper et al., 2010):

$$\hat{g}_{t|t} = \rho_g \hat{g}_{t-1} + \vartheta_{t|t} \tag{20}$$

where $|\rho_g| < 1$ and $\vartheta_{t|t} \sim iid(0, \sigma^2_\vartheta)$.

The fiscal balance rule holds:

$$\hat{g}_{t|t} = \widehat{tax}_{t|t} \tag{21}$$

Finally, the aggregate resource constraint holds true:

$$y_{t|t} = I_{t|t} + c_{t|t} + g_{t|t} \qquad (22)$$

where $I_{t|t}$ and $c_{t|t}$ label, respectively, investment and household consumption (the two components of aggregate demand expressed by the dynamic IS curve).

Although it may be a simplification compared to their real Data Generating Process (DGP), the assumption that macroeconomic shocks follow an AR(1) process is very common in the literature (Smets & Wouters, 2003; Blanchard & Galì, 2007; de Jesus et al., 2020), as it effectively balances the need to account for the persistence mechanisms of exogenous shocks with the need to avoid overly complex specifications that could make the entire model mathematically intractable. Furthermore, specific empirical analyses support the validity of this assumption for keynote estimated DSGEs (Peersman & Straub, 2006).

The model presented in this section is solved under both full and asymmetric information schemes. The full information solution is identified by solving the system defined by equations (1)-(22) where $j_{t|t} = j_t$ and $j_{t|t}$ labels the generic model's variable.

## 4. The full information solution

The full information solution of this model is determined by assuming that the central bank and the government disclose all data not included in the original information set of private agents (households and firms) without communication ambiguity or forecasting errors. Therefore, $v_{t|t} = v_t = 0$ which means that the private agents' information set coincides with that of the public sector.

### *Proposition 1*

*After switching off the exogenous shifters for the Blanchard-Khan test, the state-space representation of the model defined by equations (1) – (22) in full information ($v_t = 0$) coincides with the standard New Keynesian block with rational expectations (i.e. dynamic IS curve, New Keynesian Phillips curve, and contemporaneous Taylor rule). In this canonical case, upon setting $\alpha_y \geq 0$, $\sigma > 0$, $k > 0$, and $0 < \beta < 1$, the equilibrium determinacy and uniqueness are ensured only by the validity of the Taylor principle ($\alpha_\pi > 1$).*

See Appendix A1 for the proof.

The reduced-form solution of full information for $a_t$ is simply given by the informative shocks about the current and future values of the fundamentals and their short-run persistence:

$$a_t = \rho_q q_{t-1} + \lambda_t + \rho_b \varepsilon_{t-1}^b + \xi_t \qquad (23)$$

According to equation (23), the full-information signal provided by the public sector is a linear combination of news shocks about the current and future values of the fundamentals, with no uninformative noise component due to opacity or estimation errors ($v_t = 0$). The signal structure implies that private agents can immediately disentangle the effects of news about the current state of the economy from those concerning future economic conditions. In particular, for $h \geq 1$, $E_t[a_{t+h}|\mathcal{F}_t] = \rho_q^h q_t + \rho_b^h \varepsilon_t^b$, so that the availability of a fully transparent informative signal allows agents to form expectations about future fundamentals without confounding contemporaneous news with news about future states.

National income positively responds to the fiscal shock, demand shock, and shock to potential output, while its responses to the nominal interest rate shock and cost-push shock to inflation are negative:

$$y_t = \frac{\alpha_\pi \bar{\pi}}{\alpha_\pi k + \alpha_y + \sigma} - \frac{\alpha_\pi \rho_\epsilon}{\alpha_\pi k + \alpha_y + \sigma}\epsilon_{t-1} - \frac{\alpha_\pi}{\alpha_\pi k + \alpha_y + \sigma}\varsigma_t - \frac{\rho_i}{\alpha_\pi k + \alpha_y + \sigma}\varepsilon_{t-1}^i$$
$$- \frac{1}{\alpha_\pi k + \alpha_y + \sigma}w_t + \frac{\sigma\delta_a\rho_q}{\alpha_\pi k + \alpha_y + \sigma}q_{t-1} + \frac{\sigma\delta_a}{\alpha_\pi k + \alpha_y + \sigma}\lambda_t$$
$$+ \frac{\sigma\delta_a\rho_b}{\alpha_\pi k + \alpha_y + \sigma}\varepsilon_{t-1}^b + \frac{\sigma\delta_a}{\alpha_\pi k + \alpha_y + \sigma}\xi_t + \frac{\sigma\delta_G\rho_g}{\alpha_\pi k + \alpha_y + \sigma}\hat{g}_{t-1}$$
$$+ \frac{\sigma\delta_G}{\alpha_\pi k + \alpha_y + \sigma}\vartheta_t + \frac{\sigma\rho_{\hat{y}}}{\alpha_\pi k + \alpha_y + \sigma}\varepsilon_{t-1}^{\hat{y}} + \frac{\sigma}{\alpha_\pi k + \alpha_y + \sigma}\eta_t$$
$$+ \frac{\rho_{\bar{y}}(\alpha_\pi k + \alpha_y)}{\alpha_\pi k + \alpha_y + \sigma}\bar{y}_{t-1} + \frac{\alpha_\pi k + \alpha_y}{\alpha_\pi k + \alpha_y + \sigma}\omega_t$$

(24)

A positive shock to potential output directly affects actual output by enhancing the overall productive capacity of the economy. The expansionary effect of the demand shock arises because, upon observing an increase in aggregate demand, firms expand production to meet higher demand. The validity of the Taylor principle precludes the resulting demand pressures from translating into inflationary pressure. The same interpretation holds for the positive effect of an expansionary fiscal shock (or government spending increase). A positive shock to the nominal interest rate expands the real interest rate, with negative consequences in terms of aggregate demand and then actual output. In principle, one would expect that a positive cost-push shock to inflation would stimulate aggregate production thanks to the reduction of the real interest rate. However, this is not the case in equation (24), as the validity of the Taylor principle implies that this positive effect of the cost-push shock to inflation is more than offset by the aggressive response of the central bank to actual inflation dictated by the Taylor rule with $\alpha_\pi > 1$. The response of GDP to the news shock about either the current or future state of the economy is strictly dependent on the sign of the shock itself. Good news ($\lambda_t > 0$, $\xi_t > 0$) improve the expectations of the private agents about the fundamentals, which stimulates aggregate demand ($\hat{y}_t$) and, consequently, aggregate output ($y_t$). On the other hand, bad news worsens economic outlooks and discourages consumption and investment, which results in a decline of national income.

The responses of the output gap (aggregate demand) to the model's shocks display the same signs, except the response to the potential output shock, which is now negative:

$$\hat{y}_t = \frac{\alpha_\pi \bar{\pi}}{\alpha_\pi k + \alpha_y + \sigma} - \frac{\alpha_\pi \rho_\epsilon}{\alpha_\pi k + \alpha_y + \sigma}\epsilon_{t-1} - \frac{\alpha_\pi}{\alpha_\pi k + \alpha_y + \sigma}\varsigma_t - \frac{\rho_i}{\alpha_\pi k + \alpha_y + \sigma}\varepsilon_{t-1}^i$$
$$- \frac{1}{\alpha_\pi k + \alpha_y + \sigma}w_t + \frac{\sigma\delta_a\rho_q}{\alpha_\pi k + \alpha_y + \sigma}q_{t-1} + \frac{\sigma\delta_a}{\alpha_\pi k + \alpha_y + \sigma}\lambda_t$$
$$+ \frac{\sigma\delta_a\rho_b}{\alpha_\pi k + \alpha_y + \sigma}\varepsilon_{t-1}^b + \frac{\sigma\delta_a}{\alpha_\pi k + \alpha_y + \sigma}\xi_t + \frac{\sigma\delta_G\rho_g}{\alpha_\pi k + \alpha_y + \sigma}\hat{g}_{t-1}$$
$$+ \frac{\sigma\delta_G}{\alpha_\pi k + \alpha_y + \sigma}\vartheta_t + \frac{\sigma\rho_{\hat{y}}}{\alpha_\pi k + \alpha_y + \sigma}\varepsilon_{t-1}^{\hat{y}} + \frac{\sigma}{\alpha_\pi k + \alpha_y + \sigma}\eta_t$$
$$- \frac{\sigma\rho_{\bar{y}}}{\alpha_\pi k + \alpha_y + \sigma}\bar{y}_{t-1} - \frac{\sigma}{\alpha_\pi k + \alpha_y + \sigma}\omega_t$$

(25)

A positive shock to the nominal interest rate deteriorates the output gap via the dynamic IS curve because it increases the real interest rate. By contrast, a positive cost-push shock to inflation, by lowering the real interest rate, would be expected to enhance the output gap. Nevertheless, the enforcement of the Taylor principle involves that the monetary policy authority raises the nominal interest rate of an amount that more than compensates for the actual inflation hike, which brings the real interest rate above the pre-shock level (with a negative net effect on the output gap). A positive shock to potential output naturally decreases the output gap because of the consequent improvement of the overall productive capacity of the economy.

A positive news shock (either about the current or future state variables) boosts both consumption and investment (the two aggregate demand components, according to the dynamic IS curve) by improving the private agents' expectations about the fundamentals.

Actual inflation in deviation from its constant target ($\hat{\pi}_t = \pi_t - \bar{\pi}$) positively responds to the fiscal and demand shocks, while its reaction to the nominal interest rate and potential output shocks is negative:

$$
\begin{aligned}
\hat{\pi}_t = {} & \frac{\alpha_\pi k \bar{\pi}}{\alpha_\pi k + \alpha_y + \sigma} + \frac{\rho_\epsilon (\alpha_y + \sigma)}{\alpha_\pi k + \alpha_y + \sigma} \epsilon_{t-1} + \frac{\alpha_y + \sigma}{\alpha_\pi k + \alpha_y + \sigma} \varsigma_t - \frac{\rho_i k}{\alpha_\pi k + \alpha_y + \sigma} \varepsilon_{t-1}^i \\
& - \frac{k}{\alpha_\pi k + \alpha_y + \sigma} w_t + \frac{\sigma \delta_a \rho_q k}{\alpha_\pi k + \alpha_y + \sigma} q_{t-1} + \frac{\sigma k \delta_a}{\alpha_\pi k + \alpha_y + \sigma} \lambda_t \\
& + \frac{\sigma k \delta_a \rho_b}{\alpha_\pi k + \alpha_y + \sigma} \varepsilon_{t-1}^b + \frac{\sigma k \delta_a}{\alpha_\pi k + \alpha_y + \sigma} \xi_t + \frac{\sigma k \delta_G \rho_g}{\alpha_\pi k + \alpha_y + \sigma} \hat{g}_{t-1} \\
& + \frac{\sigma k \delta_G}{\alpha_\pi k + \alpha_y + \sigma} \vartheta_t + \frac{\sigma k \rho_{\hat{y}}}{\alpha_\pi k + \alpha_y + \sigma} \varepsilon_{t-1}^{\hat{y}} + \frac{\sigma k}{\alpha_\pi k + \alpha_y + \sigma} \eta_t \\
& - \frac{\sigma k \rho_{\bar{y}}}{\alpha_\pi k + \alpha_y + \sigma} \bar{y}_{t-1} - \frac{\sigma k}{\alpha_\pi k + \alpha_y + \sigma} \omega_t
\end{aligned}
\tag{26}
$$

A positive shock to the nominal interest rate contracts the output gap through the dynamic IS curve, which in turn leads to the reduction of actual inflation by the New Keynesian Phillips curve. A positive fiscal shock or demand shock contributes positively to the aggregate demand (via the dynamic IS curve), which leads to inflationary pressure. Of course, upon observing the growth of the output gap, the central bank responds by raising the nominal interest rate, but this policy reaction does not completely eliminate the effect of the fiscal and demand shock on inflation. A positive shock to potential output downsizes the output gap, which causes the decline of actual inflation via the New Keynesian Phillips curve.

A positive news shock (either to the current or future values of the fundamentals) improves private sector expectations and therefore lifts actual inflation by the New Keynesian Phillips curve.

The central bank reacts to the shocks hitting government spending, aggregate demand, and inflation by increasing the nominal interest rate in deviation, while the response of the policy rate in deviation to the potential output shock is negative:

$$
\begin{aligned}
\hat{\imath}_t = {} & \frac{\alpha_\pi \bar{\pi} (\alpha_\pi k + \alpha_y)}{\alpha_\pi k + \alpha_y + \sigma} + \frac{\rho_\epsilon \alpha_\pi \sigma}{\alpha_\pi k + \alpha_y + \sigma} \epsilon_{t-1} + \frac{\alpha_\pi \sigma}{\alpha_\pi k + \alpha_y + \sigma} \varsigma_t + \frac{\rho_i \sigma}{\alpha_\pi k + \alpha_y + \sigma} \varepsilon_{t-1}^i \\
& + \frac{\sigma}{\alpha_\pi k + \alpha_y + \sigma} w_t + \frac{\sigma \delta_a \rho_q (\alpha_\pi k + \alpha_y)}{\alpha_\pi k + \alpha_y + \sigma} q_{t-1} + \frac{\sigma \delta_a (\alpha_\pi k + \alpha_y)}{\alpha_\pi k + \alpha_y + \sigma} \lambda_t \\
& + \frac{\sigma \delta_a \rho_b (\alpha_\pi k + \alpha_y)}{\alpha_\pi k + \alpha_y + \sigma} \varepsilon_{t-1}^b + \frac{\sigma \delta_a (\alpha_\pi k + \alpha_y)}{\alpha_\pi k + \alpha_y + \sigma} \xi_t \\
& + \frac{\sigma \delta_G \rho_g (\alpha_\pi k + \alpha_y)}{\alpha_\pi k + \alpha_y + \sigma} \hat{g}_{t-1} + \frac{\sigma \delta_G (\alpha_\pi k + \alpha_y)}{\alpha_\pi k + \alpha_y + \sigma} \vartheta_t + \frac{\sigma \rho_{\hat{y}} (\alpha_\pi k + \alpha_y)}{\alpha_\pi k + \alpha_y + \sigma} \varepsilon_{t-1}^{\hat{y}} \\
& + \frac{\sigma (\alpha_\pi k + \alpha_y)}{\alpha_\pi k + \alpha_y + \sigma} \eta_t - \frac{\sigma \rho_{\bar{y}} (\alpha_\pi k + \alpha_y)}{\alpha_\pi k + \alpha_y + \sigma} \bar{y}_{t-1} - \frac{\sigma (\alpha_\pi k + \alpha_y)}{\alpha_\pi k + \alpha_y + \sigma} \omega_t
\end{aligned}
\tag{27}
$$

In equation (27), the signs of the parameters associated with the cost-push shock to inflation and shock to potential output are consistent with the policy coefficients of the contemporaneous Taylor rule: the central bank ensures equilibrium determinacy and uniqueness by increasing its policy rate in the face of an upsurge in the actual inflation and/or output gap eventually triggered by positive shocks to these two variables. Since a positive shock to potential output shrinks the output gap, the Taylor rule prescribes a decrease in the nominal interest rate.

Equation (27) implies that, under contemporaneous policy setting, the central bank responds to positive news shocks by raising the nominal interest rate. While this reaction dampens the expansionary effects of news on inflation and the output gap and accelerates convergence to the steady state, it plays no role in guaranteeing equilibrium determinacy and uniqueness, which continue to hinge exclusively on the Taylor principle (See Appendix A1)

The job-finding intensity in deviation ($\hat{f}_t = f_t - \bar{f}_t$) negatively reacts to the cost-push shock to inflation, shock to the nominal interest rate, and shock to potential output, while its responses to the other kinds of shocks are ambiguous and depend on the coefficients of the structural representation of the model:

$$
\begin{aligned}
\hat{f}_t = {} & \frac{\alpha_\pi \bar{\pi}[\phi_y - \phi_r(\alpha_\pi k + \alpha_y)]}{\alpha_\pi k + \alpha_y + \sigma} + \rho_f \hat{f}_{t-1} - \frac{\alpha_\pi \rho_\epsilon(\phi_y + \sigma\phi_r)}{\alpha_\pi k + \alpha_y + \sigma} \epsilon_{t-1} - \frac{\alpha_\pi(\phi_y + \sigma\phi_r)}{\alpha_\pi k + \alpha_y + \sigma} \varsigma_t \\
& - \frac{\rho_i(\phi_y + \sigma\phi_r)}{\alpha_\pi k + \alpha_y + \sigma} \varepsilon_{t-1}^i - \frac{\phi_y + \sigma\phi_r}{\alpha_\pi k + \alpha_y + \sigma} w_t \\
& + \frac{\sigma\delta_a \rho_q[\phi_y - \phi_r(\alpha_\pi k + \alpha_y)]}{\alpha_\pi k + \alpha_y + \sigma} q_{t-1} + \frac{\sigma\delta_a[\phi_y - \phi_r(\alpha_\pi k + \alpha_y)]}{\alpha_\pi k + \alpha_y + \sigma} \lambda_t \\
& + \frac{\sigma\delta_a \rho_b[\phi_y - \phi_r(\alpha_\pi k + \alpha_y)]}{\alpha_\pi k + \alpha_y + \sigma} \varepsilon_{t-1}^b + \frac{\sigma\delta_a[\phi_y - \phi_r(\alpha_\pi k + \alpha_y)]}{\alpha_\pi k + \alpha_y + \sigma} \xi_t \\
& + \frac{\sigma\delta_G \rho_g[\phi_y - \phi_r(\alpha_\pi k + \alpha_y)]}{\alpha_\pi k + \alpha_y + \sigma} \hat{g}_{t-1} + \frac{\sigma\delta_G[\phi_y - \phi_r(\alpha_\pi k + \alpha_y)]}{\alpha_\pi k + \alpha_y + \sigma} \vartheta_t \\
& + \frac{\sigma\rho_{\hat{y}}[\phi_y - \phi_r(\alpha_\pi k + \alpha_y)]}{\alpha_\pi k + \alpha_y + \sigma} \varepsilon_{t-1}^{\hat{y}} + \frac{\sigma[\phi_y - \phi_r(\alpha_\pi k + \alpha_y)]}{\alpha_\pi k + \alpha_y + \sigma} \eta_t \\
& - \frac{\sigma\rho_{\bar{y}}[\phi_y + \phi_r\sigma - \phi_r\rho_{\hat{y}}(\alpha_\pi k + \alpha_y + \sigma)]}{\alpha_\pi k + \alpha_y + \sigma} \bar{y}_{t-1} - \frac{\sigma(\phi_y + \phi_r\sigma)}{\alpha_\pi k + \alpha_y + \sigma} \omega_t \\
& + \varepsilon_t^f
\end{aligned}
\tag{28}
$$

As discussed when commenting on equation (24) above, under a Taylor-active policy ($\alpha_\pi > 1$) with $\alpha_y > 0$, both the cost-push shock to inflation and shock to the nominal interest rate deteriorate actual output and output gap, which naturally entails a decline in the job-finding intensity in deviation.

Similarly, the shock to potential output downsizes the output gap. Thus, the markup-adjusted marginal product is reduced, and firms find it convenient to decrease their hiring inflows.

The net effect of a news shock to either the current or future values of the fundamentals is determined by the tension between the demand effect and the monetary policy reaction effect. The former is expressed by $\sigma\delta_a\phi_y > 0$, and it reflects the positive impact of the informative shock on the output gap (aggregate demand). The second, instead, is the negative effect of the central bank's reaction to the increase in the output gap $\left(-\phi_r\delta_a(\alpha_\pi k + \alpha_y) < 0\right)$. More precisely, when observing that the news shock has implied the growth of the aggregate demand, the central bank helps stabilize dynamics by raising the nominal interest rate. As a consequence, the output gap declines and the job-finding intensity in deviation decreases.

A contemporary shock to potential output reduces the job-finding intensity in deviation $\left(-\frac{\sigma(\phi_y + \phi_r\sigma)}{\alpha_\pi k + \alpha_y + \sigma} < 0\right)$ because it lowers the output gap. Quantifying the impact of the persistence effect of the same shock $\left(-\frac{\sigma\rho_{\bar{y}}[\phi_y + \phi_r\sigma - \phi_r\rho_{\hat{y}}(\alpha_\pi k + \alpha_y + \sigma)]}{\alpha_\pi k + \alpha_y + \sigma}\right)$ is more complex, as it is necessary to disentangle three different components. The first is given by the demand channel $\left(-\sigma \rho_{\bar{y}} \phi_y < 0\right)$ and points out that the propagation of the shock to potential output leads to a persistent deterioration of the output gap, with negative consequences for the job-finding in deviation intensity. The second is the substitution effect $\left(-\phi_r \rho_{\bar{y}} \sigma^2 < 0\right)$, which indicates that the job-finding intensity in deviation is negatively affected by the change in the real interest rate induced by the persistent shock to potential output through the dynamic IS curve. The latter experiences a downward shift whose magnitude depends on the intertemporal elasticity of substitution $\frac{1}{\sigma}$: after the potential output shock, the real interest becomes higher, which decreases aggregate demand (output gap). As a result, the job-finding intensity in deviation decreases. The third is the monetary policy feedback effect $\left[\sigma \phi_r \rho_{\bar{y}}^2 (\alpha_\pi k + \alpha_y + \sigma)\right] > 0$: when the central bank sees that the persistent shock to potential output is reducing the output gap, it cuts the nominal interest rate to help restore equilibrium determinacy and uniqueness.

The job-separation intensity in deviation ($\hat{s}_t = s_t - \bar{s}_t$) positively reacts to the cost-push shock to inflation and shock to the nominal interest rate, while its response to all the other types of shocks is ambiguous:

$$
\begin{aligned}
\hat{s}_t = & -\frac{\alpha_\pi \bar{\pi}\left[\psi_y - \psi_r(\alpha_\pi k + \alpha_y)\right]}{\alpha_\pi k + \alpha_y + \sigma} + \rho_s \hat{s}_{t-1} + \frac{\alpha_\pi \rho_\epsilon (\psi_y + \sigma \psi_r)}{\alpha_\pi k + \alpha_y + \sigma} \epsilon_{t-1} \\
& + \frac{\alpha_\pi(\psi_y + \sigma \psi_r)}{\alpha_\pi k + \alpha_y + \sigma} \varsigma_t + \frac{\rho_i(\psi_y + \sigma \psi_r)}{\alpha_\pi k + \alpha_y + \sigma} \varepsilon_{t-1}^i + \frac{\psi_y + \sigma \psi_r}{\alpha_\pi k + \alpha_y + \sigma} w_t \\
& - \frac{\sigma \delta_a \rho_q \left[\psi_y - \psi_r(\alpha_\pi k + \alpha_y)\right]}{\alpha_\pi k + \alpha_y + \sigma} q_{t-1} - \frac{\sigma \delta_a \left[\psi_y - \psi_r(\alpha_\pi k + \alpha_y)\right]}{\alpha_\pi k + \alpha_y + \sigma} \lambda_t \\
& - \frac{\sigma \delta_a \rho_b \left[\psi_y - \psi_r(\alpha_\pi k + \alpha_y)\right]}{\alpha_\pi k + \alpha_y + \sigma} \varepsilon_{t-1}^b - \frac{\sigma \delta_a \left[\psi_y - \psi_r(\alpha_\pi k + \alpha_y)\right]}{\alpha_\pi k + \alpha_y + \sigma} \xi_t \\
& - \frac{\sigma \delta_G \rho_g \left[\psi_y - \psi_r(\alpha_\pi k + \alpha_y)\right]}{\alpha_\pi k + \alpha_y + \sigma} \hat{g}_{t-1} - \frac{\sigma \delta_G \left[\psi_y - \psi_r(\alpha_\pi k + \alpha_y)\right]}{\alpha_\pi k + \alpha_y + \sigma} \vartheta_t \\
& - \frac{\sigma \rho_{\hat{y}} \left[\psi_y - \psi_r(\alpha_\pi k + \alpha_y)\right]}{\alpha_\pi k + \alpha_y + \sigma} \varepsilon_{t-1}^{\hat{y}} - \frac{\sigma \left[\psi_y - \psi_r(\alpha_\pi k + \alpha_y)\right]}{\alpha_\pi k + \alpha_y + \sigma} \eta_t \\
& + \frac{\sigma \rho_{\bar{y}} \left[\psi_y + \psi_r \sigma - \rho_{\bar{y}} \psi_r(\alpha_\pi k + \alpha_y + \sigma)\right]}{\alpha_\pi k + \alpha_y + \sigma} \bar{y}_{t-1} + \frac{\sigma(\psi_y + \sigma \psi_r)}{\alpha_\pi k + \alpha_y + \sigma} \omega_t \\
& + \varepsilon_t^s
\end{aligned} \tag{29}
$$

Job-separation intensity in deviation grows in response to the nominal interest rate shock and cost push shock to inflation.

The signs of the effects of the informative shocks are dictated by the complex interaction between the demand effect ($\psi_y$) and the policy reaction effect $\left(\psi_r(\alpha_\pi k + \alpha_y)\right)$. For $\psi_y > \psi_r(\alpha_\pi k + \alpha_y)$, positive news shocks about the current and future values of the fundamentals reduce job-separation intensity in deviation. The interpretation of this result is as follows: the positive news shocks increase both aggregate demand ($\hat{y}_t$) and actual inflation ($\hat{\pi}_t$), and thus job-separation intensity in deviation declines. Upon observing the output gap's rise and actual inflation's growth, the central bank raises its nominal interest rate ($\hat{\imath}_t$), which reduces both aggregate demand and aggregate output. As a result, job-separation intensity in deviation goes down. Hence, the net effect of a news shock (either about the current or future state of the economy) depends on the difference between the demand effect and the monetary policy feedback effect: it is positive when the demand effect dominates the monetary policy feedback effect, and negative otherwise.

The same mathematical condition $\left(\psi_y > \psi_r(\alpha_\pi k + \alpha_y)\right)$ establishes the impact of a demand shock on job-separation intensity in deviation, and its economic meaning coincides with that discussed above for the generic news shock.

As concerns the role of potential output, it is necessary to distinguish between the contemporaneous effect of the shock to this variable (the coefficient of equation (29) associated with $\omega_t$) and its short-run persistence (the coefficient of equation (29) associated with $\bar{y}_{t-1}$).

The effect of a positive contemporaneous shock to potential output is always positive, as it instantaneously reduces the output gap.

A positive shock to potential output occurred at time $t-1$ (namely, the persistence effect of $\omega_t$), reduces job-separation intensity in deviation at time $t$ only if $\sigma\rho_{\bar{y}}\{\psi_y + \psi_r[(1-\rho_{\bar{y}})\sigma - \rho_{\bar{y}}(\alpha_\pi k + \alpha_y)]\} > 0$. This mathematical condition clarifies that the persistence effect of the potential output shock on job-separation intensity in deviation is made of three parts. The first, which is always positive, is $\sigma\rho_{\bar{y}}\psi_y > 0$, and it represents the demand channel: the potential output shock contracts the output gap, which increases the markup-adjusted marginal product and spurs job-separations. The second (the substitution effect), which is even positive $\left(\sigma^2\rho_{\bar{y}}\psi_r(1-\rho_{\bar{y}})\right) > 0$, is given by the effect of the persistent potential output shock occurred at time $t-1$ ($\rho_{\bar{y}}$) on the real interest rate, whose size depends on the slope of the dynamic IS curve $\left(\frac{1}{\sigma}\right)$. In practice, the persistence effect of the potential output shock occurred at time $t-1$ downward shifts the dynamic IS curve by $\frac{1}{\sigma}$, which changes the real interest rate and increases the costs of consuming and investing today. Then, aggregate demand (output gap) declines and the job-separation intensity lifts.

The third, instead, is negative $\left(-\sigma\psi_r\rho_{\bar{y}}^2(\alpha_\pi k + \alpha_y) < 0\right)$, and it is again the feedback effect of monetary policy: the output gap's contraction induces the central bank to cut the nominal interest rate, which revives aggregate demand.

The first difference of government debt in deviation positively responds to all the model's shocks except those hitting potential output (whose reduced-form coefficient is negative):

$$
\begin{aligned}
\Delta\hat{d}_t ={}& \frac{\alpha_\pi \bar{d}\bar{\pi}(\alpha_\pi k + \alpha_y)}{\alpha_\pi k + \alpha_y + \sigma} + \frac{\bar{d}\rho_\epsilon \alpha_\pi \sigma}{\alpha_\pi k + \alpha_y + \sigma}\epsilon_{t-1} + \frac{\bar{d}\alpha_\pi \sigma}{\alpha_\pi k + \alpha_y + \sigma}\varsigma_t + \frac{\bar{d}\rho_i \sigma}{\alpha_\pi k + \alpha_y + \sigma}\varepsilon_{t-1}^i \\
&+ \frac{\sigma\bar{d}}{\alpha_\pi k + \alpha_y + \sigma}w_t + \frac{\sigma\delta_a\rho_q\bar{d}(\alpha_\pi k + \alpha_y)}{\alpha_\pi k + \alpha_y + \sigma}q_{t-1} + \frac{\sigma\delta_a\bar{d}(\alpha_\pi k + \alpha_y)}{\alpha_\pi k + \alpha_y + \sigma}\lambda_t \\
&+ \frac{\sigma\delta_a\rho_b\bar{d}(\alpha_\pi k + \alpha_y)}{\alpha_\pi k + \alpha_y + \sigma}\varepsilon_{t-1}^b + \frac{\sigma\delta_a\bar{d}(\alpha_\pi k + \alpha_y)}{\alpha_\pi k + \alpha_y + \sigma}\xi_t \\
&+ \frac{\sigma\delta_G\rho_g\bar{d}(\alpha_\pi k + \alpha_y)}{\alpha_\pi k + \alpha_y + \sigma}\hat{g}_{t-1} + \frac{\sigma\delta_G\bar{d}(\alpha_\pi k + \alpha_y)}{\alpha_\pi k + \alpha_y + \sigma}\vartheta_t \\
&+ \frac{\sigma\rho_{\hat{y}}\bar{d}(\alpha_\pi k + \alpha_y)}{\alpha_\pi k + \alpha_y + \sigma}\varepsilon_{t-1}^{\hat{y}} + \frac{\sigma\bar{d}(\alpha_\pi k + \alpha_y)}{\alpha_\pi k + \alpha_y + \sigma}\eta_t - \frac{\sigma\rho_{\bar{y}}\bar{d}(\alpha_\pi k + \alpha_y)}{\alpha_\pi k + \alpha_y + \sigma}\bar{y}_{t-1} \\
&- \frac{\sigma\bar{d}(\alpha_\pi k + \alpha_y)}{\alpha_\pi k + \alpha_y + \sigma}\omega_t
\end{aligned}
\tag{30}
$$

When the Taylor principle holds, a positive shock to the nominal interest rate makes its real counterpart higher, and the interest payments on the outstanding debt become more expansive for the public sector. Because of the enforcement of the Taylor principle, a positive cost-push shock to inflation raises the real interest rate while deteriorating GDP and (consequently) tax revenue. This harms the sustainability of public debt. A positive shock to the output gap results in an excess of demand that entails an increase in the actual inflation rate, which, in turn, induces the central bank to raise the nominal interest rate. Hence, the cost of the debt service (real interest rate) grows.

By contrast, a positive shock to potential output decreases government debt, because it contributes to narrowing the output gap. Therefore, this type of shock removes the excess of demand in the economy, together with the related inflationary effects, and finally decreases the cost of the debt service.

Interestingly, positive news shocks about the current and future values of the fundamentals worsen the debt sustainability. The reason for this unintuitive effect is that, because of the fiscal balance rule, there is no extra tax revenue when actual output grows, while the output gap (namely, aggregate demand) rises. Again, this generates inflationary pressures and induces the central bank to respond with an increase in the real interest rate, which lessens government debt's sustainability.

Private agents' current expectations about future job-separation intensity in deviation at time $t + h$ responds to both exogenous shocks (and their respective persistence terms) exactly like the actual job-separation intensity in deviation at time $t$:

$$
\begin{aligned}
E_t[\hat{s}_{t+h}^{FI}|\mathcal{F}_t] = & -\frac{\alpha_\pi \bar{\pi}[\psi_y - \psi_r(\alpha_\pi k + \alpha_y)]}{\alpha_\pi k + \alpha_y + \sigma} + \rho_s^h \hat{s}_{t-1} + \rho_s^{h-1} \varepsilon_t^s \\
& + \frac{\alpha_\pi \rho_e^h(\psi_y + \sigma \psi_r)}{\alpha_\pi k + \alpha_y + \sigma} \epsilon_{t-1} + \frac{\alpha_\pi \rho_e^{h-1}(\psi_y + \sigma \psi_r)}{\alpha_\pi k + \alpha_y + \sigma} \varsigma_t \\
& + \frac{\rho_i^h(\psi_y + \sigma \psi_r)}{\alpha_\pi k + \alpha_y + \sigma} \varepsilon_{t-1}^i + \frac{\rho_i^{h-1}(\psi_y + \sigma \psi_r)}{\alpha_\pi k + \alpha_y + \sigma} w_t \\
& - \frac{\sigma \delta_a \rho_q^h[\psi_y - \psi_r(\alpha_\pi k + \alpha_y)]}{\alpha_\pi k + \alpha_y + \sigma} q_{t-1} - \frac{\sigma \delta_a \rho_q^{h-1}[\psi_y - \psi_r(\alpha_\pi k + \alpha_y)]}{\alpha_\pi k + \alpha_y + \sigma} \lambda_t \\
& - \frac{\sigma \delta_a \rho_b^h[\psi_y - \psi_r(\alpha_\pi k + \alpha_y)]}{\alpha_\pi k + \alpha_y + \sigma} \varepsilon_{t-1}^b - \frac{\sigma \delta_a \rho_b^{h-1}[\psi_y - \psi_r(\alpha_\pi k + \alpha_y)]}{\alpha_\pi k + \alpha_y + \sigma} \xi_t \\
& - \frac{\sigma \delta_G \rho_g^h[\psi_y - \psi_r(\alpha_\pi k + \alpha_y)]}{\alpha_\pi k + \alpha_y + \sigma} \hat{g}_{t-1} \\
& - \frac{\sigma \delta_G \rho_g^{h-1}[\psi_y - \psi_r(\alpha_\pi k + \alpha_y)]}{\alpha_\pi k + \alpha_y + \sigma} \vartheta_t - \frac{\sigma \rho_{\hat{y}}^h[\psi_y - \psi_r(\alpha_\pi k + \alpha_y)]}{\alpha_\pi k + \alpha_y + \sigma} \varepsilon_{t-1}^{\hat{y}} \\
& - \frac{\sigma \rho_{\hat{y}}^{h-1}[\psi_y - \psi_r(\alpha_\pi k + \alpha_y)]}{\alpha_\pi k + \alpha_y + \sigma} \eta_t \\
& + \frac{\sigma \rho_{\bar{y}}^h[\psi_y + \psi_r \sigma - \rho_{\bar{y}} \psi_r(\alpha_\pi k + \alpha_y + \sigma)]}{\alpha_\pi k + \alpha_y + \sigma} \bar{y}_{t-1} \\
& + \frac{\sigma \rho_{\bar{y}}^{h-1}[\psi_y + \psi_r \sigma - \rho_{\bar{y}} \psi_r(\alpha_\pi k + \alpha_y + \sigma)]}{\alpha_\pi k + \alpha_y + \sigma} \omega_t
\end{aligned}
\tag{31}
$$

The interpretations of the theoretical coefficients in equation (31) are perfectly aligned with those in equation (29), even if, for $h \geq 1$, $\varepsilon_t^s = \varsigma_t = w_t = \lambda_t = \xi_t = \vartheta_t = \eta_t = \omega_t = 0$, which definitely simplifies this expression

Again, the net impact of a news shock about the current or future values of the fundamentals on the expected job-separation intensity depends on the prevalence of its demand effect over the policy reaction of the central bank.

Equation (31) represents the starting base to develop a comprehensive modeling of job insecurity as an ad-hoc survival-hazard function.

## 5. Asymmetric information analysis

After studying the characterization of the equilibrium solution in full information, it is useful to delve into the alternative scenario in which the informative signal of the public sector is noisy $v_t \neq 0$.

### Proposition 2

*Upon switching off the exogenous shifters for the Blanchard-Khan test and assuming that the structural parameters of the model are in their standard domain ($\alpha_y \geq 0$, $\sigma > 0$, $k > 0$, and $0 < \beta < 1$), the necessary and sufficient condition for equilibrium determinacy and uniqueness in the asymmetric information case ($v_t \neq 0$) is the strict validity of the Taylor principle ($\alpha_\pi > 1$).*

See Appendix A1 for the proof.

The model's solution is unique, but it cannot be expressed as an elementary closed-form in which each endogenous variable is a finite polynomial of the fundamental shocks. This important difference with the case analysed in Proposition 1 is due to the fact that asymmetric information introduces into the model a filtered component whose Data Generating Process (DGP) is an infinite moving average process ($MA(\infty)$).

Hence, the closed-form solution is given by the linear law of motion of the state vector obtained by the Kalman filtration technique, and the theoretical equation of the expected job-separation intensity in deviation with asymmetric information and active Taylor principle at time $t + h$ is delivered by:

$$E_t[\hat{s}_{t+h}^{AI}|\mathcal{H}_t] = c^\top A^h \tilde{x}_t \tag{32}$$

where $c$ is the loading vector, $A$ is the transition matrix, and $\tilde{x}_t$ is the vector of filtered endogenous variables.

The unique balanced growth path of the model under full information and a binding Taylor principle differs from that obtained under asymmetric information, even when the Taylor principle holds. The reason is that the latter incorporates a filtered component of the noisy term $v_t$, estimated by private agents, which is absent under full information. Consequently, the expected job-separation intensity in deviation of full information differs from its asymmetric information counterpart ($E_t[\hat{s}_{t+h}^{FI}|\mathcal{F}_t] \neq E_t[\hat{s}_{t+h}^{AI}|\mathcal{H}_t]$).

It is important to underscore that, in order to make the Kalman filter applicable, it is necessary to impose at least one sufficient identification restriction ensuring that the private information set $\mathcal{H}_t$ satisfies the observability condition of the model's state-space representation. Such sufficient but not necessary restrictions may be as follows:

1) The persistence of the news shocks about the current state of the economy differs from that of the informative shocks about the future values of the fundamentals ($\rho_q \neq \rho_b$).
2) The variances of the two informative shocks substantially depart from each other.
3) The correlations of the two news shocks with the economic fundamentals are distinct.
4) Private agents filter a combination of the two informative shocks, not each of them separately.

The comparison between Proposition 1 and Proposition 2 reveals that full institutional transparency entails a welfare benefit to the extent that it removes the informative noise (whose filtered component otherwise contributes to determining the equilibrium dynamics of the endogenous variables). However, institutional transparency alone is not sufficient to guarantee the determination and uniqueness of the equilibrium. To this end, it must be accompanied by the satisfaction of the Taylor principle.

### First Corollary to Proposition 1 and Proposition 2

*When the Taylor principle is enforced (regardless of the information hierarchy) and by leveraging the expression of the expectations formulated at the current time regarding the job-separation intensity in deviation at time $t + h$, the subjective expectation of losing one's job between time $t$ and time $t +$*

*h at least once can be expressed (in continuous time) by the following standard survival-hazard mapping:*

$$JI_t^{[0,h],\Re} = 1 - E_t\left[e^{-\int_t^{t+H} \hat{s}_\tau^{\Re} d\tau} \mid \mathcal{Q}_t\right] \tag{33}$$

*where $\Re \in \{FI, AI\}$, $\mathcal{Q}_t \in \{\mathcal{F}_t, \mathcal{H}_t\}$, and $\tau \in [t, t+H]$*

*Upon defining $\Delta$ the sampling step of the model, equation (33), using some approximations, job-insecurity equation (33) can be rearranged as follows (See Appendix A2):*

$$JI_t^{[0,h],\Re} \approx 1 - e^{(-\Lambda_t(h)^{\Re})} \tag{34}$$

*where $\Lambda_t(h)^{\Re}$ is the Riemann sum with step $\Delta$:*

$$\Lambda_t(h)^{\Re} \approx \Delta \sum_{h=1}^{H} E_t[\hat{s}_{t+h}^{\Re} \mid \mathcal{Q}_t] \tag{35}$$

*and $H = \frac{\Delta}{h}$.*

Equations (34)-(35) yield a closed-form solution for job-insecurity that can be practically applied to the data after estimating the equilibrium expected job-separation intensity in deviation defined either as in Proposition 1 or as in Proposition 2, but they have two important limitations. The first is that they account for the "mean-only" component of job-insecurity (namely, the mean of the process) without paying attention to the variance of $\hat{s}_t^{\Re}$ between $t+1$ and $t+H$ conditional to the informative signal $\left(Var(\hat{s}_{t+1}^{\Re}, \ldots, \hat{s}_{t+H}^{\Re} \mid a_t)\right)$. The latter is a measure of uncertainty and neglecting it can lead to overstating job insecurity (especially for large $H$ and high volatility of $\hat{s}_{t+h}^{\Re}$). The second problem is that expressions (34) and (35) are naturally affected by a certain approximation error. Indeed, one can demonstrate that, for a plausible range of values of $H$, $\Delta$, and $\rho_s$, the approximation errors of expressions (34) and (35) are negligible (See Appendix A3) but eliminating them remains desirable. These two problems can be addressed by *i)* assuming that $\hat{s}_\tau^{\Re}$ is a piecewise-constant continuous time job-separation intensity as in the discrete hazard models; *ii)* defining $\hat{S}_{N,t}^{\Re} = -\int_t^{t+H} \hat{s}_\tau^{\Re} d\tau = \Delta \sum_{h=1}^{H} E_t[\hat{s}_{t+h}^{\Re} \mid \mathcal{Q}_t] = \Lambda_t(h)^{\Re}$; *iii)* making the additional hypotheses that the linearized model's shocks are normally distributed and the cumulated job-separation intensity between time $t$ and time $t+h$ $(\hat{S}_{N,t}^{\Re})$ is conditionally Gaussian (which does not compromise the validity of Proposition 1 and Proposition 2); and *iv)* finally applying the following second-order risk correction (See Appendix A3):

$$JI_t^{[0,h],\Re} = 1 - e^{\left(-\mu_{S,t}^{\Re} + \frac{1}{2}\sigma_{S,t}^{2,\Re}\right)} \tag{36}$$

where:

- $\mu_{S,t}^{\Re} \equiv \Lambda_t(h)^{\Re}$;
- $\sigma_{S,t}^{2,\Re} = Var_t(\hat{S}_{N,t}^{\Re} \mid \mathcal{Q}_t)$;
- $\hat{S}_{N,t}^{\Re} \mid \Theta_t \sim N(\mu_{S,t}^{\Re}, \sigma_{S,t}^{2,\Re})$;
- $\Theta_t$ is the vector of exogenous shocks.

The hypothesis that $\hat{s}_\tau^{\Re}$ is piecewise-constant (which is itself sufficient to make expressions (34) and (35) hold as equalities) is consistent with discrete-time hazard models and can be rationalized by the fact that hiring and firing decisions are typically tied to contractual and institutional arrangements that operate at fixed horizons. In a representative-agent setting, it is therefore reasonable to approximate the aggregate separation intensity within $[t, t+1]$ by a constant intensity determined at time $t$.

In equation (36), when the expectations formulated today about cumulated job-separation intensity in deviation between time $t$ and time $t+h$ meets its steady-state level ($E_t\big[\hat{S}_{N,t}^{\Re}|\mathcal{Q}_t\big]=0$), current job insecurity is zero ($JI_t^{[0,h],\Re}=0$), namely, the risk of job loss perceived between time $t$ and time $t+h$ is neither higher nor lower than its long-run benchmark. When instead cumulated job-separation intensity in deviation expected at time $t+h$ is above its steady-state level ($E_t\big[\hat{S}_{N,t}^{\Re}|\mathcal{Q}_t\big]>0$), job insecurity is positive ($JI_t^{[0,h],\Re}>0$). Finally, when cumulated job-separation intensity expected at time $t+h$ lies below its steady-state counterpart ($E_t\big[\hat{S}_{N,t}^{\Re}|\mathcal{Q}_t\big]<0$), job insecurity is negative ($JI_t^{[0,h],\Re}<0$), namely, it can be interpreted as a job security macroeconomic indicator (lower-than-normal, perceived risk of job loss).

***Second Corollary to Proposition 1 and Proposition 2***

*Define the variance of the cumulated job-separation intensity in deviation between time $t+1$ and time $t+H$ ($\hat{S}_{N,t+h}^{\Re}$) as:*

$$Var_t\big(\hat{S}_{N,t+1}^{\Re}|\mathcal{Q}_t, \ldots, \hat{S}_{N,t+H}^{\Re}|\mathcal{Q}_t\big) = \mathcal{D}_t \Sigma_{H,t} \mathcal{D}_t^{\mathsf{T}} \tag{37}$$

*where*:

-  *$\mathcal{D}_t \in \mathbb{R}^{H\times m}$ is the dynamic weight matrix such that the relationship between cumulated job-separation intensity in deviation and the generic shock to variable $j$ at time $t+h$ with $1\leq h < H$ (namely, $\mathcal{E}_{t+h}^{j}$) is mapped by:*

$$\begin{bmatrix} \hat{S}_{N,t+1}^{\Re}|\mathcal{Q}_t \\ . \\ . \\ . \\ \hat{S}_{N,t+H}^{\Re}|\mathcal{Q}_t \end{bmatrix} = \mathcal{D}_t \begin{bmatrix} \mathcal{E}_{t+1}^{j} \\ . \\ . \\ . \\ \mathcal{E}_{t+H}^{j} \end{bmatrix}$$

-  *$\Sigma_{H,t} = diag(\Sigma_{t+1}, \ldots, \Sigma_{t+H}) \in \mathbb{R}^{mH\times mH}$ with $\Sigma_{t+h} = Var_t(\mathcal{E}_{t+h})$ being diagonal if and only if the model's shocks are mutually independent and $\mathcal{E}_{t+h} \in \mathbb{R}^{H\times m}$ (the assumption of mutual independence is adopted for expositional clarity and can be relaxed without affecting the qualitative results)*;
-  *$i = (1, \ldots, 1)^{\mathsf{T}} \in \mathbb{R}^{H}$;*
-  *$\Delta > 0$.*

*Set also:*

$$\mu_{S,t}^{\Re} = \Delta i^{\mathsf{T}} E_t \begin{bmatrix} \hat{S}_{N,t+1}^{\Re}|\mathcal{Q}_t \\ . \\ . \\ . \\ \hat{S}_{N,t+H}^{\Re}|\mathcal{Q}_t \end{bmatrix}, \qquad \sigma_{S,t}^{2,\Re} = \Delta^2 i^{\mathsf{T}} \mathcal{D}_t \Sigma_{H,t} \mathcal{D}_t^{\mathsf{T}} i$$

*such that the definition of job-insecurity $JI_t^{[0,h],\Re}$ is consistent with the second-order representation given in equation (36).*

*Finally, let:*

$$j_{t+1} = \mathcal{V}j_t + \mathcal{B}e_j\mathcal{E}_{t+1}^{j}$$

*be the transition equation of the j-th endogenous variable of the reduced-form linearized representation of the model, where:*

- $\mathcal{V} \in \mathbb{R}^{H \times H}$ *is the transition matrix;*
- $\mathcal{B} \in \mathbb{R}^{H \times m}$ *is the shock-impact matrix;*
- $e_j \in \mathbb{R}^m$ *is the fixed canonical vector that selects the shock to variable j;*

*Under the Taylor principle ($\alpha_\pi > 1$) and for $\mathfrak{R} \in \{FI, AI\}$, the impact of the generic exogenous shock to variable j at time t is nonlinear, and it is given by:*

$$\frac{\partial JI_t^{[0,h],\mathfrak{R}}}{\partial \varepsilon_t^j} = e^{\left(-\mu_{S,t}^{\mathfrak{R}} + \frac{1}{2}\sigma_{S,t}^{2,\mathfrak{R}}\right)} \left[ \Delta \mathcal{K}^\top \left( \sum_{h=0}^{H-1} \mathcal{V}^h \right) \mathcal{B} e_j \right.$$
$$\left. - \frac{1}{2} \Delta^2 \iota^\top \left( \frac{d\mathcal{D}_t}{d\varepsilon_t^j} \Sigma_{H,t} \mathcal{D}_t^\top + \mathcal{D}_t \Sigma_{H,t} \frac{d\mathcal{D}_t^\top}{d\varepsilon_t^j} + \mathcal{D}_t \frac{d\Sigma_{H,t}}{d\varepsilon_t^j} \mathcal{D}_t^\top \right) \iota \right]$$
(38)

*where:*

- $\mu_{S,t}^{\mathfrak{R}} = \Delta \mathcal{K}^\top (\sum_{h=0}^{H-1} \mathcal{V}^h) \mathcal{B} e_j$ *is the "mean effect" of the generic shock (which is explained by the dynamic multipliers);*
- $\sigma_{S,t}^{2,\mathfrak{R}} = \Delta^2 \iota^\top \left( \frac{d\mathcal{D}_t}{d\varepsilon_t^j} \Sigma_{H,t} \mathcal{D}_t^\top + \mathcal{D}_t \Sigma_{H,t} \frac{d\mathcal{D}_t^\top}{d\varepsilon_t^j} + \mathcal{D}_t \frac{d\Sigma_{H,t}}{d\varepsilon_t^j} \mathcal{D}_t^\top \right) \iota$ *is the "variance effect" (stochastic volatility) of the generic shock, which is a generalization of the case in which the variance effect becomes time-varying;*
- $\mathcal{K} \in \mathbb{R}^{1 \times H}$ *is loading vector of the reduced form representation of the New Keynesian model.*

In equation (38), the second-order risk correction ($\sigma_{S,t}^{2,\mathfrak{R}} \neq 0$) can be implemented only upon including stochastic volatility into the expressions of the model's exogenous shocks, which entails rewriting each shock as a system formed by two equations (Kim et al., 1998):

$$y_{t|t} = \rho_y y_{t-1} + \sigma_{\mathcal{E},t|t}^2 \mathcal{E}_{t|t}$$
(39)

and:

$$h_{\mathcal{E},t|t} \equiv \log(\sigma_{\mathcal{E},t|t}^2) = \mu_{\mathcal{E}} + \chi_{\mathcal{E}} [h_{\mathcal{E},t-1} - \mu_{\mathcal{E}}] + \sigma_{h,\mathcal{E}} x_{\mathcal{E},t|t}$$
(40)

where:

- $y_t \in \left( q_{t|t}, \varepsilon_{t|t}^b, \bar{y}_{t|t}, \varepsilon_{t|t}^{\hat{y}}, \epsilon_{t|t}, \varepsilon_{t|t}^i, \bar{f}_{t|t}, \bar{s}_{t|t} \right)$;
- $\rho_y \in \left( \rho_q, \rho_b, \rho_{\bar{y}}, \rho_{\hat{y}}, \rho_e, \rho_i, \rho_f, \rho_s \right)$;
- $\mathcal{E}_t \in \left( \lambda_{t|t}, \xi_{t|t}, \omega_{t|t}, \eta_{t|t}, \varsigma_{t|t}, w_{t|t}, \varepsilon_{t|t}^f, \varepsilon_{t|t}^s \right)$;
- $x_{\mathcal{E},t} \in \left( x_{\lambda,t}, x_{\xi,t}, x_{\omega,t}, x_{\eta,t}, x_{\varsigma,t}, x_{w,t}, x_{f,t}, x_{s,t} \right)$;
- $h_{\mathcal{E},t|t} \in \left( h_{\lambda,t|t}, h_{\xi,t|t}, h_{\omega,t|t}, h_{\eta,t|t}, h_{\varsigma,t|t}, h_{w,t|t}, h_{f,t|t}, h_{s,t|t} \right)$;
- $\sigma_{\mathcal{E},t|t}^2 \in \left( \sigma_{\lambda,t|t}^2, \sigma_{\xi,t|t}^2, \sigma_{\omega,t|t}^2, \sigma_{\eta,t|t}^2, \sigma_{\varsigma,t|t}^2, \sigma_{w,t|t}^2, \sigma_{f,t|t}^2, \sigma_{s,t|t}^2 \right)$;
- $\sigma_{h,\mathcal{E}} \in \left( \sigma_{h,\lambda}, \sigma_{h,\xi}, \sigma_{h,\omega}, \sigma_{h,\eta}, \sigma_{h,\varsigma}, \sigma_{h,w}, \sigma_{h,f}, \sigma_{h,s} \right)$;
- $\mu_{\mathcal{E}} \in \left( \mu_\lambda, \mu_\xi, \mu_\omega, \mu_\eta, \mu_\varsigma, \mu_w, \mu_f, \mu_s \right)$;
- $\chi_{\mathcal{E}} \in \left( \chi_\lambda, \chi_\xi, \chi_\omega, \chi_\eta, \chi_\varsigma, \chi_w, \chi_f, \chi_s \right)$;
- $\mathcal{E}_{t|t} \sim N(0,1)$ *and* $x_{\mathcal{E},t|t} \sim N(0,1)$.

Proposition 1, Proposition 2, and the related corollaries are fully robust to assumptions (39) – (40), because the stochastic volatility hypothesis affects only the conditional variances, not the dynamic of the linear means. The unique change in the balanced growth paths equations of the endogenous variables (23) – (30) is that now the theoretical coefficients associated with the shocks comprised in vector $\mathcal{b}_t$ are pre-multiplied by the respective stochastic variance terms (the elements of vector $\sigma_{\mathcal{b},t|t}^2$). The economic interpretation of the equilibrium solutions in the setting providing for full information and Taylor principle is not sensitive to this slight change.

In equation (38), the impact of the generic shock is always nonlinear, but its sign is endogenous to the estimated parameters of the reduced-form (namely, the multipliers of the IS-NKPC-Taylor rule block that map the relationship between the informative shocks and the job-separation intensity).

Finally, it is important to clarify that the stochastic volatility block above is introduced solely to discipline the second-order risk correction and is not required for the existence or determinacy of the equilibrium.

### Proposition 3

*The asymmetric information version ($\nu_t \neq 0$) of the New-Keynesian model formed by equations (1) – (22) admits a belief-driven (sunspot) equilibrium anchored to the innovation of the informative signal ($\Gamma_t = a_t - E_{t-1}[a_t|\mathcal{H}_{t-1}]$) if and only if the following conditions hold:*

1) *The Taylor principle does not hold ($\alpha_\pi \leq 1$) and the other parameters are in their standard domain ($\alpha_y \geq 0$, $\sigma > 0$, $k > 0$ and $0 < \beta < 1$);*

2) *The extrinsic driver to which the expectations are anchored, namely, $\mathbb{Z}_t = \rho_{\mathbb{Z}}\mathbb{Z}_{t-1} + \Gamma_t$, is a stationary AR(1) process ($|\rho_{\mathbb{Z}}| < 1$) whose innovation $\Gamma_t$ is orthogonal to all the current and past values of the shocks to the fundamentals:*

$$\Gamma_t \perp \left( a_\tau, q_\tau, \varepsilon_\tau^v, \lambda_\tau, \xi_\tau, \bar{y}_\tau, \bar{f}_\tau, \bar{s}_\tau, \omega_\tau, \varepsilon_\tau^{\hat{y}}, \eta_\tau, \bar{r}_\tau, \varepsilon_\tau^f, \varepsilon_\tau^s, \epsilon_\tau, \varsigma_\tau, \varepsilon_\tau^i, w_\tau, \vartheta_\tau \right), \qquad \tau \leq t$$

3) *Given the companion of the dynamic IS-NKPC-Taylor rule block:*

$$j_{t+1} = \mathcal{M}j_t + F\mathcal{s}\mathbb{Z}_t + \mathcal{N}\varepsilon_{t+1}^j$$

*whose dynamic belief matrix is:*

$$\widetilde{\mathcal{M}} = \begin{bmatrix} A & F\mathcal{s} \\ 0 & \rho_{\mathbb{Z}} \end{bmatrix}$$

*for $\alpha_\pi \leq 1$, there exists a value of $\rho_{\mathbb{Z}} \in (-1, +1)$ such that the spectrum of $\widetilde{\mathcal{M}}$ has exactly two stable eigenvalues (the same number as the model's forward-looking variables) and thus the compatibility condition below is satisfied:*

$$\left( 1 + \frac{\alpha_y}{\sigma} - \rho_{\mathbb{Z}} \right)(1 - \beta\rho_{\mathbb{Z}}) - k \left( \frac{\alpha_\pi}{\sigma} - \frac{\rho_{\mathbb{Z}}}{\sigma} \right) = 0$$

*Otherwise, the Blanchard-Khan criterion is already satisfied with $\mathbb{Z}_t = \mathcal{s} = 0$, and then the equilibrium is unique (fundamental) and coincides with the canonical one (Proposition 2).*

See Appendix A4 for the proof.

The second assumption underlying Proposition 3 requires that private agents' actual expectations about the forward-looking variables of the IS-NKPC block (inflation and output gap) are affected by $\mathbb{Z}_t$, such that they deviate from their canonical rational expectations representation (namely, $E_t[\hat{\pi}_{t|t}|\mathcal{H}_t]^{act} = E_t[\hat{\pi}_{t|t}|\mathcal{H}_t]^{RE} + \mathcal{s}_\pi\mathbb{Z}_t$ and $E_t[\hat{y}_{t|t}|\mathcal{H}_t]^{act} = E_t[\hat{y}_{t|t}|\mathcal{H}_t]^{RE} + \mathcal{s}_y\mathbb{Z}_t$). This does not amount to relaxing the rational expectations hypothesis. Rather, it amounts to allowing for belief-driven (sunspot) equilibria that are compatible with the Blanchard–Kahn conditions under a violation of the Taylor principle. In this setting, $\mathbb{z}_t$ acts as an extrinsic driver selecting a particular equilibrium within this class. Importantly, these equilibria are not E-stable and therefore are not learnable under adaptive learning, so that they are not selected by plausible expectation-formation dynamics.

Proposition 3 does not invalidate the job insecurity expression delivered by equation (36) under rational expectations, but one needs to assume that *i*) the sunspot process $\Gamma_t$ (and then $\mathbb{Z}_t$) is a normal shock and *ii*) the model's linearized version remains linear-gaussian, such that $\hat{S}_{N,t}^{\mathfrak{R}}$ is conditionally normal.

### Corollary to Proposition 3

*Under the hypotheses of Proposition 3, actual expectations about cumulated job-separation intensity in deviation between time $t$ and time $t + h$ formulated at time $t$ are given by the sum of their fundamental and non-fundamental components:*

$$E_t\big[\hat{S}_{N,t+h}^{AI}|\mathcal{H}_t\big]^{act} = E_t\big[\hat{S}_{N,t+h}^{AI}|\mathcal{H}_t\big]^{RE} + F_s\rho_{\mathbb{Z},s}^h\mathbb{Z}_t \qquad (41)$$

*Job insecurity is the same as that in equation (36) for $\mathfrak{R} = AI$ with*:

$$\mu_{S,t}^{AI} = \Delta\sum_{h=1}^{H} E_t\big[\hat{S}_{N,t+h}^{AI}|\mathcal{H}_t\big]^{RE} + \frac{\Delta F_s\rho_{\mathbb{Z},s}\big(1-\rho_{\mathbb{Z},s}^H\big)}{\big(1-\rho_{\mathbb{Z},s}\big)}\mathbb{Z}_t \qquad (42)$$

*and*:

$$\sigma_{S,t}^{2,AI} = \Delta^2 F_s^2\left[\frac{\rho_{\mathbb{Z},s}\big(1-\rho_{\mathbb{Z},s}^H\big)}{\big(1-\rho_{\mathbb{Z},s}\big)}\right]^2 Var_t(\mathbb{Z}_t) \qquad (43)$$

*where $Var_t(\mathbb{Z}_t)$ is conditionally predetermined.*

*Nevertheless, the impact of sunspot-belief shock $\mathbb{Z}_t$ on $JI_t^{[0,h],AI}$ does not depend on the second-order effects of the same shock conditional on the stochastic volatility path, but exclusively on its first order effects (because the conditional variance term $\sigma_{S,t}^{2,AI}$ is independent of $\mathbb{Z}_t$ under the maintained assumptions)*:

$$\frac{\partial JI_t^{[0,h],AI}}{\partial\mathbb{Z}_t} = \frac{\Delta F_s\rho_{\mathbb{Z},s}\big(1-\rho_{\mathbb{Z},s}^H\big)}{\big(1-\rho_{\mathbb{Z},s}\big)}e^{\big(-\mu_{S,t}^{AI}+\frac{1}{2}\sigma_{S,t}^{2,AI}\big)} \qquad (44)$$

This irrelevance result strictly hinges on the standard assumption that stochastic volatility is independent from the sunspot belief shock. In principle, the second-order effect becomes relevant under the alternative assumption that $\log\big(\sigma_{\hat{\sigma},t|t}^2\big)$ is related to $\mathbb{Z}_t$, but this would contrast with the hypotheses of Proposition 3.

### Proposition 4

*Given the state-space representation of the model and upon denoting the vector of latent fundamentals and news shocks with $\mathbb{X}_t$, it is possible to demonstrate that, for $\rho_b \neq \rho_q$, the minimum private agents' information set for Kalman filtering of $\mathbb{X}_t$ is $\mathcal{H}_t^{min} = \hbar(\{a_\tau\}_{\tau\leq t})$, where $\mathcal{H}_t^{min} \subseteq \mathcal{F}_t$ and $\hbar(.)$ is the $\sigma$-algebra generated by the informative signal $a_t$.*

Proposition 4 (whose proof is in Appendix A5) indicates that, if all the other conditions for their existence are satisfied, the asymmetric information equilibria explored in Proposition 2 and Proposition 3 are robust to restricting the private agents' information set to the history of informative signal $a_t$. In other words, all the theoretical predictions of the model still hold when relaxing the assumption that the actual inflation rate, nominal interest rate, real interest rate, real stock of government debt, and past values of the endogenous variables are observable to all agents.

When private agents are unable to directly observe the policy rate and need to project it, the right-hand side of the dynamic IS curve contains the expected value of this variable instead of its current realization. Consequently, there is a certain time mismatch between the time in which the central bank makes its communication about the nominal interest rate and the time in which private agents make their intertemporal choice (Sorge and Vota, 2025). According to Proposition 4, this slight modification to the dynamic IS curve does not alter the equilibrium path of asymmetric information, notwithstanding the presence of sunspot belief shocks resulting from the filtration of the informative signal. Hence, one can conclude that, at least in the asymmetric information scenario, the timing of the intertemporal choice is irrelevant both in determining the dynamics of the fundamentals and job insecurity.

This result reflects a standard property of linear-Gaussian RE models: when the policy rule is known and the policy instrument conveys no additional information on the structural shocks, delays in its observation do not affect the agents' filtered expectations nor the resulting equilibrium.

## 6. Learnability of Proposition 1 and Proposition 2

Since the conventional assumption of rational expectations (RE) is often seen as barely realistic (Dombeck, 2021), it is appropriate to evaluate the sensitivity of the theoretical results presented in Proposition 1, Proposition 2 and the related corollaries to the alternative hypothesis of Adaptive Learning (AL).

To this end, consider the state-space representation of the model:

$$\Phi_0 E_t[\gamma_{t+1}^{\Re}|\mathcal{Q}_t] = \Phi_1 \gamma_t + \Phi_2 \Xi_t \tag{45}$$

where $\mathcal{Q}_t \in (\mathcal{F}_t, \mathcal{H}_t)$ and $\Re \in (FI, AI)$, while:

$$\gamma_t = \begin{bmatrix} \hat{\pi}_{t+1} \\ \hat{y}_{t+1} \end{bmatrix}, \qquad \Phi_0 = \begin{bmatrix} \beta & 0 \\ -\frac{1}{\sigma} & -1 \end{bmatrix}, \qquad \Phi_1 = \begin{bmatrix} 1 & k \\ \frac{\alpha_\pi}{\sigma} & \frac{\alpha_y + \sigma}{\sigma} \end{bmatrix}, \qquad \Phi_2 = \begin{bmatrix} 1 & 0 & 0 & 0 & 0 \\ 0 & \frac{1}{\sigma} & -\frac{1}{\sigma} & -\delta_a & -\delta_G \end{bmatrix}, \qquad \Xi_t = \begin{bmatrix} \varepsilon_t^\pi \\ \varepsilon_t^i \\ \bar{r}_t \\ a_t \\ \hat{g}_t \end{bmatrix}$$

Assume that, under linear learning with decreasing gain, private agents form their expectations according to the following Perceived Law of Motion (PLM):

$$E_t[\gamma_{t+1}^{\Re}|\mathcal{Q}_t] = K_1 \gamma_t + K_2 \Xi_t \tag{46}$$

where $K_1 \in \mathbb{R}^{2 \times 2}$ and $K_2 \in \mathbb{R}^{2 \times 5}$ are the matrices containing the coefficients updating via Recursive Least Squares (or similar algorithms).

By replacing the PLM (46) in the state-space representation (45), one obtains the Actual Law of Motion (ALM) of the model with Adaptive Learning:

$$(\Phi_0 K_1 - \Phi_1)\gamma_t + (\Phi_0 K_2 - \Phi_2)\Xi_t = 0 \tag{47}$$

Equation (47) needs to be satisfied for each $(\gamma_t, \Xi_t)$, which requires finding out the fixed point of the linear system on its left-hand side. Thus, the learning process is mapped by the equations below (T-map):

$$T(K_1) = \Phi_0^{-1} \Phi_1, \qquad T(K_2) = \Phi_0^{-1} \Phi_2 \tag{48}$$

whose fixed point $(K_1^*, K_2^*)$ is given by:

$$K_1^* = \Phi_0^{-1}\Phi_1, \qquad K_2^* = \Phi_0^{-1}\Phi_2 \tag{49}$$

and this is exactly the same Rational Expectations Equilibrium (REE) solution calculated by directly solving the system.

Following Evans and Honkapohja (2009), the parameters of the T-map continuously update according to the ordinary differential equations below:

$$\dot{K}_1 = T(K_1) - K_1 = \Phi_0^{-1}\Phi_1 - K_1 \tag{50}$$

and:

$$\dot{K}_2 = T(K_2) - K_2 = \Phi_0^{-1}\Phi_2 - K_2 \tag{51}$$

Equations (50) and (51) are affine, and they are nonlinearly independent from $K_1$ and $K_2$. Hence, upon rewriting both equations in terms of deviation from the fixed point $(K_1^*, K_2^*)$, namely, $\widetilde{K}_1 = K_1 - K_1^*$ and $\widetilde{K}_2 = K_2 - K_2^*$, they become, respectively:

$$\dot{\widetilde{K}}_1 = -\left(\widetilde{K}_1\right) \tag{52}$$

and:

$$\dot{\widetilde{K}}_2 = -\left(\widetilde{K}_2\right) \tag{53}$$

The solutions of equations (52) and (53) are, respectively, $\widetilde{K}_1(t) = e^{-t}\widetilde{K}_1(0)$ and $\widetilde{K}_2(t) = e^{-t}\widetilde{K}_2(0)$, which implies that matrices $K_1$ and $K_2$ globally and exponentially converge to $K_1^*$ and $K_2^*$ for any initial condition $\left((K_1, K_2) \to (K_1^*, K_2^*)\right)$.

Since all the eigenvalues of the system formed by the ordinary differential equations (52) and (53) are equal to $-1$ and T-map (48) is constant, the REE equilibrium $(K_1^*, K_2^*)$ is also E-stable with no further parametric restrictions other than the invertibility of $\Phi_0$:

$$\det(\Phi_0) = 0 \Longleftrightarrow \sigma + \alpha_y - \alpha_\pi k \neq 0 \tag{54}$$

which is the usual condition imposed to deduce the state-space representation of the model.

The conclusion of the sensitivity analysis above is that Proposition 1 and Proposition 2 are robust to Adaptive Learning, in the sense that the equilibrium selected under RE matches the limit of the learning dynamics (E-stability).

The intuition behind the mathematical results of this section is that the dynamic IS-NKPC-Taylor rule block includes only forward-looking variables, and, consequently, the private sector needs to learn about the elements of only two constant matrices (namely, $K_1$ and $K_2$). However, the mapping of the model is independent from $K_1$ and $K_2$, which makes the automatic convergence of the AL equilibrium to its REE counterpart feasible.

The implication of this finding is that job insecurity is learnable in the sense of E-stability, as private agents' expectations of adaptive learning converge to the RE benchmark. As a result, if the conditions of Propositions 1 and 2 hold: *i*) job insecurity remains solely driven by short-run persistent fundamental shocks; *ii*) the mathematical expression of job insecurity given in equations (34)-(36) remains valid; and *iii*) the responses of job insecurity to fundamental shocks coincide with those obtained under rational expectations.

## 7. Non-learnability of Proposition 3

Following the same analytical approach as that in the previous section, it is possible to prove that Proposition 3 is not robust to the adaptive-learning hypothesis.

Consider again the state-space representation of the model by equation (45) with $\Re = AI$ and $Q = \mathcal{H}_t$. In addition, define matrices $A_1 = \Phi_0^{-1}\Phi_1$ and $A_2 = \Phi_0^{-1}\Phi_2$ and vector of exogenous shifters $\mathbb{s}_t = [\varepsilon_t^\pi \quad \varepsilon_t^i \quad \bar{r}_t \quad q_t \quad \varepsilon_t^b \quad \nu_t \quad \hat{g}_t \quad \mathbb{Z}_t]^\top \in \mathbb{R}^8$ whose law of motion is $\mathbb{s}_t = \mathbb{F}_s \mathbb{s}_{t-1} + \mathbb{v}_t$, where $\mathbb{v}_t$ is a martingale difference sequence $(E_t[\mathbb{v}_t|\mathcal{H}_{t-1}] = 0$ and $Var[\mathbb{v}_t|\mathcal{H}_{t-1}] = \sigma_{\mathbb{v},t}^2)$ and $\mathbb{F}_s = diag(\rho_\varepsilon, \rho_i, 0, \rho_q, \rho_b, 0, \rho_g, \rho_{\mathbb{Z}})$. The nexus between exogenous variables and shifters is expressed by selection matrix $\mathbb{W}$:

$$\Xi_t = \mathbb{W}\mathbb{s}_t, \qquad \mathbb{W} = \begin{bmatrix} 1 & 0 & 0 & 0 & 0 & 0 & 0 & 0 \\ 0 & 1 & 0 & 0 & 0 & 0 & 0 & 0 \\ 0 & 0 & 1 & 0 & 0 & 0 & 0 & 0 \\ 0 & 0 & 0 & 1 & 1 & 1 & 0 & 0 \\ 0 & 0 & 0 & 0 & 0 & 0 & 1 & 0 \end{bmatrix} \in \mathbb{R}^{5\times 8} \tag{55}$$

The expectations formation process of private agents is dictated by the following PLM:

$$\gamma_t = \alpha_0 + B^{(e)}\mathbb{s}_t \tag{56}$$

where $B^{(e)} \in R^{2\times 8}$.

By replacing equation (56) in equation (45), one obtains the corresponding ALM:

$$\gamma_t = A_1\alpha_0 + T(B^{(e)}) = A_1\alpha_0 + (A_1 B^{(e)}\mathbb{F}_s + A_2\mathbb{W})\mathbb{s}_t \tag{57}$$

where $T(B^{(e)}) \in \mathbb{R}^{2\times 5}$ is the model's T-map.

In order to assess the model's E-stability, one needs to vectorize $b = \text{vec}(B^{(e)})$ as below:

$$\text{vec}[T(B^{(e)})] = (\mathbb{F}_s \otimes A_1)\,\text{vec}(B^{(e)}) + \text{vec}(A_2\mathbb{W}) \tag{58}$$

Upon defining the updating map set by ALM (57) as $\mathcal{T}(b) = \text{vec}(A_1 B^{(e)}\mathbb{F}_s + A_2\mathbb{W})$, the decreasing gain limit of the ordinary differential equation is $\dot{b} = \mathcal{T}(b) - b$, which yields the Jacobian associated with T-map (57):

$$\mathcal{J} = \frac{\partial \text{vec}\left(T(B^{(e)})\right)}{\partial B^{(e)}} = \mathbb{F}_s^\top \otimes A_1 - I \tag{59}$$

which is clearly independent from the term $A_2\mathbb{W}$.

The E-stability condition is that, in equation (59), all the eigenvalues of $\mathbb{F}_s^\top \otimes A_1$ have real part less than 1 (namely, $\varrho(\mathbb{F}_s^\top \otimes A_1) = \varrho(\mathbb{F}_s)\varrho(A_2) < 1$). Since, under the conditions of Proposition 3, $\alpha_\pi \leq 1 \Rightarrow \varrho(A_2) \geq 1$, then, $\varrho(\mathbb{F}_s)\varrho(A_2) \geq 1 \forall A_2\mathbb{W}$, and the E-stability condition is not satisfied.

This result has powerful implications, as it proves that, if the assumptions of Proposition 3 hold, *i)* job insecurity of AL with decreasing gain does not converge toward a unique equilibrium path, because $E_t[\hat{S}_{N,t+h}^{AI}|\mathcal{H}_t]$ does not move toward the RE benchmark; *ii)* keeping constant the fundamentals, job insecurity of AL with decreasing gain can change as a result of either extrinsic driver $\mathbb{Z}_t$ or initial learning conditions; *iii)* even in the absence of fundamental innovations, the sunspot belief shocks and changes in the initial learning conditions can generate endogenous volatility and persistence in the job insecurity of AL with decreasing gain; *iv)* the mathematical expressions provided by equations (34) and (36) are not reliable, because they do not account for the belief-driven changes in the job insecurity of AL with decreasing gain; *v)* only the validity of the Taylor principle ($\alpha_\pi > 1$) restores equilibrium determinacy in AL with decreasing gain and makes job insecurity purely fundamentals-driven.

In other words, non-learnability implies that agents cannot anchor their expectations through observed outcomes, so that job insecurity becomes inherently history-dependent and sensitive to belief shocks, even in the absence of fundamental innovations

## 8. The New Keynesian model with cohort heterogeneity

The literature on job insecurity has been paying increasing attention to specific categories of vulnerable workers like the mature ones (Näswall & De Witte, 2003; Lain et al., 2019; Bertolini et al., 2024; Errichiello & Falavigna, 2024), which are conventionally defined as workers aged 50 and over. For this reason, it is worthwhile to extend the baseline NK model proposed in Section 3 to the case of cohort heterogeneity.

Upon labeling under 50 and over 50 workers respectively as $YO$ and $OL$, $CO \in (YO, OL)$ is the reference cohort. Define cohort-specific laws of motion for the unemployment rate ($u_t^{CO} \in (u_t^{YO}, u_t^{OL})$), job-separation intensity ($s_t^{CO} \in (s_t^{YO}, s_t^{OL})$), and job-finding intensity ($f_t^{CO} \in (f_t^{YO}, f_t^{OL})$). Then, aggregate them as below:

$$\begin{cases} u_{t|t} = \phi_u^{YO} u_{t|t}^{YO} + \phi_u^{OL} u_{t|t}^{OL} \\ f_{t|t} = \phi_f^{YO} f_{t|t}^{YO} + \phi_f^{OL} f_{t|t}^{OL} \\ s_{t|t} = \phi_s^{YO} s_{t|t}^{YO} + \phi_s^{OL} s_{t|t}^{OL} \end{cases} \tag{60}$$

where:
- $\phi_f^{YO} + \phi_f^{OL} = 1$;
- $\phi_s^{YO} + \phi_s^{OL} = 1$:
- $\phi_u^{YO} + \phi_u^{OL} = 1$.

Here, $\phi_u^{YO}$ and $\phi_u^{OL}$ are the employment/demographic shares of the under 50 and over 50 cohorts, which are constant in steady-state. $\phi_f^{YO}$ and $\phi_f^{OL}$, conversely, are the shares of hirings and separations of cohort $CO$ relative to the total which are constant in steady-state.

In order to incorporate cohort-specific nominal and real rigidities (including wage markups), the New Keynesian Phillips Curve slack parameter is expressed using the steady-state shares, namely, $\bar{k} = \phi_k^{YO} k^{YO} + \phi_k^{OL} k^{OL}$, where $\phi_k^{YO}$ and $\phi_k^{OL}$ are functions of the steady-state markup/rigidity terms.

The equations dictating other relationships between the endogenous variables are the same as the baseline version of the model.

Cohort heterogeneity does not change the claims of Proposition 1: the dynamic IS-NKPC-Taylor rule block remains forward looking, and, for $\alpha_\pi > 1$, the model's equilibrium solution is still unique. The only difference with the baseline version of the model in Section 3 is that now the equations of the balanced growth paths of the endogenous variables depend on cohort-specific slack parameter $\bar{k}$ instead of $k$ as well as the other cohort-specific coefficients in system (60), namely, $\phi_f^{CO}, \phi_s^{CO}, \phi_u^{CO}, \rho_f^{CO}, \phi_y^{CO}, \phi_r^{CO}, \rho_s^{CO}, \psi_y^{CO}$, and $\psi_r^{CO}$. This change carries over to the mean-only and second-order, reduced-form expressions of aggregate job insecurity: they become a weighted average of their cohort-specific counterparts whose weights are linear combinations of parameters $\phi_f^{CO}, \phi_s^{CO}, \phi_u^{CO}, \rho_f^{CO}, \phi_y^{CO}, \phi_r^{CO}, \rho_s^{CO}, \psi_y^{CO}$, and $\psi_r^{CO}$. As a consequence, the responses of job insecurity to the shocks to the endogenous variables are affected by the same set of cohort-specific parameters.

Analogous robustness considerations apply to Proposition 2 under cohort-heterogeneity. Of course, reduced-form job insecurity does not admit a closed-form mathematical expression even with cohort heterogeneity (as in the baseline model).

Both Proposition 1 and Proposition 2 with cohort heterogeneity are learnable under AL with decreasing gain.

Finally, Proposition 3 is qualitatively unaltered by cohort heterogeneity: the compatibility condition for the existence of the belief-driven equilibrium depends on the spectrum of the dynamic IS-NKPC-Taylor rule block and the state persistence. However, the cohort-specific parameters can either expand or restrict the region of parameters satisfying the compatibility condition, because it introduces additional persistence terms ($\bar{r}_{t|t}^{CO}$, $\bar{s}_{t|t}^{CO}$) with the related aggregation weights (which are again combinations of $\phi_f^{CO}$, $\phi_s^{CO}$, $\phi_u^{CO}$, $\rho_f^{CO}$, $\phi_y^{CO}$, $\phi_r^{CO}$, $\rho_s^{CO}$, $\psi_y^{CO}$, and $\psi_r^{CO}$). The non-learnability of Proposition 3 under AL with decreasing gain is confirmed: Although the T-map satisfies the local E-stability condition when $\alpha_\pi \leq 1$, the equilibrium remains non-learnable because it is driven by extrinsic belief shocks rather than fundamentals. Thus, the implications of non-learnability such as endogenous persistence/volatility, dependence on the initial conditions, and sunspot transfer to aggregate job insecurity defined as the weighted average of its cohort-specific counterparts (namely, $JI_t^{AI} = \phi_{JI}^{YO} JI_t^{AI,YO} + \phi_{JI}^{OL} JI_t^{AI,OL}$).

## 9. Validation strategy of Proposition 1 and Proposition 2

The theoretical predictions in Proposition 1 and Proposition 2 are empirically validated by means of the usual Simulated Moment Matching method (also known as the Indirect Inference approach), which is quite popular in the empirical DSGE literature (Gourieroux et al., 1993; Smith, 1993; Ruge-Murcia, 2012).

The parameters of the structural representation of the model (IS-NKPC-Taylor rule and labour market blocks) are selected on the basis of the established literature (Clarida et al., 1999; Petrongolo & Pissarides, 2001; Christiano et al., 2005; Shimer, 2005; Smets & Wouters, 2007; Fujita & Ramey, 2009; Gertler & Trigari, 2009). Only the policy coefficient of the Taylor rule associated with the actual inflation rate is exogenously set to 1.15 and 0.85 to ensure the compliance of the simulation exercises with the hypotheses of Proposition 1 and Proposition 2.

Following Christiano et al. (2005) and Smets & Wouters (2007), the autoregressive coefficients of the 9 AR(1) shocks with Gaussian innovations are treated as estimating parameters, and their initial values lie in the interval conventionally computed in the DSGE models (0.70 - 0.95). In order to avoid the proliferation of the estimating parameters, stochastic volatility is not enclosed in the shocks' autoregressive modelling.

The estimating statistical moments are the most relevant ones for the purpose of analysing the dynamics of the key endogenous variables of the model, namely, mean, standard deviation, global autocorrelation at lag 1, and cross-correlations of actual inflation in deviation, output gap, nominal interest rate in deviation, job-finding intensity in deviation, job-separation intensity in deviation, unemployment rate, and government debt in deviation.

The starting values of the estimating moments replicate those observed in the Italian economy in the sample period 2004Q1 - 2025Q2. Of course, the simulated time series (variables and shocks) are defined in the same sample period.

The parameters of interest are simulated by the Differential Evolution algorithm for global optimization of real-valued functions. This global optimization estimator remains robust even when computing several parameters from relatively short time series.

The weighting matrix used in the algorithm (after properly standardizing the moments) is the identity one. The reason for this choice is that the standard optimal alternative, that is the inverse of the estimated variance-covariance matrix of the moments, can suffer from ill-conditioning (namely, the algorithm can find out several local minima that hinder the convergence of the theoretical vector of moments to its actual counterpart) when there are many estimating moments and a short sample period. The identity weight matrix is less asymptotically efficient than the optimal one, but (subject to no significant model misspecification) it grants the consistency of the SMM estimator.

The convergence criterion of the algorithm is tailored to a tolerance level which is sufficiently high to achieve the stability of the estimating parameters.

The fact that the number of estimating moments (18) exceeds the number of estimating parameters (9) ensures the algorithm's overidentification.

The values of the initialized parameters of the structural representation of the model together with the estimated autoregressive coefficients are reported in Table 1:

**Table 1** Coefficients used for calibration purposes under Proposition 1 and Proposition 2

| (1)<br>Parameter | (2)<br>Value under Proposition 1 | (3)<br>Value under Proposition 2 |
|:---:|:---:|:---:|
| $\beta$ | 0.99 | 0.99 |
| $k$ | 0.10 | 0.10 |
| $\sigma$ | 1.00 | 1.00 |
| $\alpha_y$ | 0.10 | 0.10 |
| $\alpha_\pi$ | 1.15 | 1.15 |
| $\phi_y$ | 0.60 | 0.60 |
| $\phi_r$ | 0.20 | 0.20 |
| $\psi_y$ | 0.60 | 0.60 |
| $\psi_r$ | 0.20 | 0.20 |
| $\delta_g$ | 0.25 | 0.25 |
| $\delta_\lambda$ | 0.35 | 0.35 |
| $\delta_\xi$ | 0.35 | 0.35 |
| $\rho_\epsilon$ | 0.78 | 0.17 |
| $\rho_g$ | 0.11 | 0.41 |
| $\rho_i$ | 0.19 | 0.93 |
| $\rho_{\hat{y}}$ | 0.60 | 0.66 |
| $\rho_{\bar{y}}$ | 0.19 | 0.40 |
| $\rho_b$ | 0.20 | 0.92 |
| $\rho_q$ | 0.24 | 0.59 |
| $\rho_f$ | 0.93 | 0.48 |
| $\rho_s$ | 0.94 | 0.93 |
| $\sigma_\lambda^2$ | 0.10 | 0.10 |
| $\sigma_\xi^2$ | 0.10 | 0.10 |
| $\sigma_\omega^2$ | 0.20 | 0.20 |
| $\sigma_\vartheta^2$ | 0.30 | 0.30 |
| $\sigma_\zeta^2$ | 0.20 | 0.20 |
| $\sigma_w^2$ | 0.10 | 0.10 |
| $\sigma_f^2$ | 0.05 | 0.05 |
| $\sigma_s^2$ | 0.05 | 0.05 |
| $\sigma_\eta^2$ | 0.20 | 0.20 |

It is important to clarify that the scope of the validation exercise is not the assessment of the model's capacity to fit the observed moments under the assumptions of Proposition 1 and Proposition 2. This would be actually impossible, because the Data Generating Processes (DGPs) of Proposition 1 (full information with Taylor principle) and Proposition 2 (asymmetric information with Taylor principle) depart from that of the Italian data observed in the sample period 2004Q1 - 2025Q2 (the AI with no Taylor principle case investigated in Proposition 3).

Thus, the aim of the simulation displayed in this section is the qualitative validation of Proposition 1 and Proposition 2 through a suitable set of parameters rather than the appraisal of the predictive capacity of the New Keynesian model.

The same simulations are replicated on data related to mature workers (over 50) with the aim of assessing the robustness of Proposition 1 and Proposition 2 to the extension of the baseline version of the model to cohort-heterogeneity.

Finally, Italy is treated as a small open economy within a monetary union: it takes the ECB policy rate as given and cannot influence area-wide monetary conditions. The policy rate is therefore exogenous to domestic macroeconomic dynamics, and the Taylor-type rule used in the model should be read as a reduced-form reaction function that summarizes the behaviour of the common monetary authority. This modelling choice is standard in country-level NK applications under EMU and does not affect the transmission mechanism generating job-insecurity dynamics (Galí and Monacelli, 2005; Christoffel, Coenen and Warne, 2008; Forni, Monteforte and Sessa, 2009). Clarifying this point is important because the empirical validation relies on Italian data while monetary policy is set at the supranational level.

## 10. A cautionary validation scheme for Proposition 3

The identification difficulties associated with sunspot belief shocks, together with the parametric restrictions required to isolate their contribution, make a direct validation of Proposition 3 through the Simulated Method of Moments (SMM) practically infeasible. In addition, as discussed above, equation (36) for job insecurity is applicable to Proposition 3 only under the assumption that the (potential) sunspot shock is normally distributed. This assumption is not fully consistent with the empirical evidence: some contributions retain Gaussian belief shocks (Dave & Sorge, 2020), whereas others document fat-tailed distributions (Kozlowski et al., 2020), discontinuous coordination patterns (Angeletos & Werning, 2006), or nonlinear belief dynamics emerging from learning (Fehr et al., 2019; Arifovic et al., 2020).

Since Proposition 3 allows for belief-driven (sunspot) components in job-insecurity dynamics, over and above fundamental shocks, the theoretical job-insecurity measure cannot be directly mapped into observable data. Any empirical counterpart therefore requires an explicit measurement layer, and the constructed series should be interpreted as a noisy indicator rather than a direct measure of the theoretical object.

For these reasons, an appropriate empirical strategy to cautionary validate Proposition 3 is to construct a survey-based index that proxies perceived job insecurity and to use it as a diagnostic time-series object to be compared with the theoretical job-insecurity measures implied by Proposition 1 (full information with the Taylor principle) and Proposition 2 (asymmetric information with the Taylor principle). To discipline the plausibility of the monetary-policy environment underlying Proposition 3, an ECB Taylor-type rule is estimated via both GMM and 2SLS over 2004Q1–2025Q2 (see Appendix A6). The instrument set includes lagged inflation and lagged output gap, as standard in forward-looking policy-rule estimation (Clarida et al., 2000), and the Brent crude-oil price as an exogenous global cost-push shifter (Favero & Rovelli, 2003; Kilian, 2009; Lubik & Schorfheide, 2004). The resulting estimates suggest an inflation-response coefficient below unity (and not statistically different from zero in the baseline specification), which is consistent with a sustained deviation from the Taylor principle over this period. Proposition 3 can therefore be treated as a plausible benchmark for organising the evidence, while the empirical exercise remains conditional on the policy-regime diagnostic and on the informational frictions featured in the model.

The survey-based index of perceived job insecurity for Italy is constructed by using the survey data on consumer confidence collected by the National Institute of Statistics (ISTAT) between 2004Q1 and 2025Q2. The respondents are asked, among the other questions, to declare whether, in the 12 months following the interview, they expect:

1) A high increase in the unemployment rate ($u_t^{HI}$);
2) A moderate increase in the unemployment rate ($u_t^{MI}$);
3) No increase in the unemployment rate ($u_t^{NI}$);
4) A low decrease in the unemployment rate ($u_t^{LD}$);

5)  A significant decrease in the unemployment rate ($u_t^{SD}$).

The database published by ISTAT reports the percentages of respondents opting for each of the five possible answers above. The weighted index formula for the index is given by:

$$\widetilde{JI}_t^{[0,12],ISTAT} = w_{HI}u_t^{HI} + w_{MI}u_t^{MI} + w_{NI}u_t^{NI} + w_{LD}u_t^{LD} + w_{SD}u_t^{SD} \tag{61}$$

where $w_{HI} = 1$, $w_{MI} = 0.5$, $w_{NI} = 0$, $w_{LD} = -0.5$, and $w_{SD} = -1$.

Since the theoretical indices $JI_t^{[0,h],FI}$ and $JI_t^{[0,h],AI}$ are estimated on job insecurity in deviation from its long-run equilibrium level, $JI_t^{[0,12],ISTAT}$ is demeaned by its trend component estimated via Ordinary Least Squares (OLS) as in the Hamilton filtering technique (2018). The resulting cyclical component is labeled with $\widetilde{JI}_t^{[0,12],ISTAT}$.

Finally, in order to make $\widetilde{JI}_t^{[0,12],ISTAT}$ comparable with $JI_t^{[0,12],FI}$ and $JI_t^{[0,12],AI}$, their respective time series are standardized (to avoid unit mismatch) and the following OLS regression model with parametric scale is estimated:

$$\widetilde{JI}_t^{[0,12],ISTAT} = \alpha_0 + \alpha_1 JI_t^{[0,12],FI} + \alpha_2 JI_t^{[0,12],AI} + \varepsilon_t^{ISTAT} \tag{62}$$

In equations (62), vertical intercept $\alpha_0$ captures the mean differences between the survey-based index of perceived job insecurity and its theoretical counterparts. The parameter scales $\alpha_1$ and $\alpha_2$, instead, express the comovement of the survey-based index of perceived job insecurity $\widetilde{JI}_t^{[0,12],ISTAT}$ with the two theoretical measures for the full information and asymmetric information cases. In particular, $\hat{\alpha}_1 = 0$ and $\hat{\alpha}_2 \neq 0$ indicate that asymmetric information substantially contributes to determine the fluctuations of the survey-based index of perceived job insecurity. On the contrary, $\hat{\alpha}_1 \neq 0$ and $\hat{\alpha}_2 = 0$ underscores that the informative signal released by the public agents has been sufficiently strong and clear to make the survey-based index of perceived job insecurity dynamics more similar to that of the full information setting rather than the asymmetric information one. The case $\alpha_1 = \alpha_2 = 0$ presumably establishes that the evolution of $\widetilde{JI}_t^{[0,12],ISTAT}$ over time is purely driven by the sunspot belief shock found in Proposition 3. Finally, $\alpha_1 \neq 0$ and $\alpha_2 \neq 0$ probably points out that the monetary policy of the European Central Bank has been sufficiently aggressive in response to inflation to make the survey-based index of perceived job insecurity close to that of the two theoretical scenarios in which the Taylor principle holds.

As concerns the validation of Proposition 3 with cohort heterogeneity, the procedure is the same as that above with the unique but relevant difference that the survey-based index of perceived job insecurity in deviation for the Italian mature workers (over 50) is computed by relying on the data available in the Survey of Health, Ageing and Retirement in Europe (SHARE) conducted between 2004 and 2022. The question used for this purpose is that related to the respondents' perspectives about their perceived job security ("*My job is not very secure*". *Would you say you strongly agree, agree, disagree, or strongly disagree?*) whose possible responses, measured on the numeric scale 1-4, are:

-   "Totally agree" (1);
-   "Agree" (2);
-    "Disagree" (3);
-   "Totally disagree" (4).

The assigned weights are $+1$ for "Totally disagree", $0.5$ for "Agree", $-0.5$ for "Disagree", and $-1$ for "Totally disagree".

The presence of stochastic volatility in the survey-based indices of perceived job insecurity (both the aggregate version and the mature workers' specification) is empirically assessed via the Markov

Chain Monte Carlo (MCMC) Bayesian methodology proposed by Kastner & Frühwirth-Schnatter (2014).

The interpretation of the estimated coefficients of equation (62) proposed above can be seen as credible only upon assuming that:

a) The parameters used to simulate job insecurity under Proposition 1 and Proposition 2 are consistent with their respective DGPs (FI with Taylor principle and AI with no Taylor principle). Otherwise, $JI_t^{[0,12],FI}$ and $JI_t^{[0,12],AI}$ are purely theoretical values useful for scenario analysis, but relating them to Z makes no sense;

b) There is no multicollinearity between $JI_t^{[0,12],FI}$ and $JI_t^{[0,12],AI}$ (which can be the case when the the variance of the uninformative noise $v_t$ is low, leading to similar patterns of job insecurity with full information and asymmetric information);

c) The relationship between $JI_t^{[0,12],FI}$, $JI_t^{[0,12],AI}$, and $\widetilde{JI}_t^{[0,12],ISTAT}$ does not include nonlinear terms which raise misspecification concerns;

d) There are no relevant measurement errors in the survey-based index of perceived job insecurity appearing on the left-hand side of equation (62) that can cause problems of dynamic contamination of the error term $\varepsilon_t^{ISTAT}$;

e) The sample size and persistence in $JI_t^{[0,12],FI}$, $JI_t^{[0,12],AI}$, and $\widetilde{JI}_t^{[0,12],ISTAT}$ are such that there are no statistical power problems.

Points b), c), and e) can be faced by carrying out the Variance Inflation Factor (VIF) and robust Ramsey RESET test and checking the dimension of the estimated coefficients and standard errors as well as the residuals correlogram. Point d) calls for selecting statistically accurate survey-based proxies of perceived job insecurity which are measured with methodologies that ensure their consistency over time. Point e) requires choosing a sufficiently long sample size and verifying that the persistence of the variables in equation (62) is moderate (which is quite probable in light of the fact that, as indicated by the theoretical model job insecurity is a linear combination of white noise shocks and AR(1) exogenous variables).

Addressing point a) is a really complex task: basically, full information is totally unrealistic while there are a few plausible examples of an economy with asymmetric information and Taylor principle (one of which is the US economy at the Great Moderation time). Hence, recovering the actual structural parameters of the DGPs of Proposition 1 and Proposition 2 is barely or totally impossible. The most logical approach to this issue is commenting on the estimation output of equation (62) by clarifying that its interpretation is viable only as long as one assumes that the structural parameters (often unobservable) on which the DGPs of Proposition 1 and Proposition 2 rely on are similar to those conventionally employed in the DSGE literature.

Although not leading to definitive conclusions, another appropriate post-estimation diagnostic check is testing for the martingale difference sequence hypothesis on the residuals of equation (62).

## 11. Data

The validation exercise has been carried out on Italian quarterly time series ranging between 2004Q1 and 2025Q2. Table 2 lists and describes each variable involved in the study to compute the empirical statistical moments of the Italian economy:

**Table 2** List of the variables used for validation purposes

| (1) Variable | (2) Unit of measure | (3) Description | (4) Source |
|---|---|---|---|
| Real GDP | Millions of Chained 2010 Euros, Seasonally Adjusted | Real Gross Domestic Product for Italy | National Accounts – GDP of Eurostat |
| Real output gap | Millions of Chained 2010 Euros, Seasonally Adjusted | Time series estimated as the cyclical component of Real GDP through the Hamilton filter (2018) | Authors' own estimation |
| Inflation rate | Growth rate of the Italian Gross Domestic Product Deflator with base date 2010Q1 = 100 | Inflation rate of the whole economy | International Financial Statistics of the International Monetary Fund |
| Unemployment rate | Percent, Seasonally adjusted | Number unemployed people aged 15 and over divided by the total labor force aged 15 and over | Main Economic Indicators of the Organization for Economic Cooperation and Development (OECD) |
| Mature workers' unemployment rate | Percent, Seasonally adjusted | Number unemployed people aged 50 and over divided by the total labor force aged 50 and over | Euro Indicators/PEEIs of Eurostat |
| Job-finding intensity | Percent, Seasonally adjusted | The quarterly transition rate (hazard/Poisson intensity) from unemployment to employment between time $t$ and time $t + 1$ conditional to being unemployed at time $t$. | Authors' elaboration on ISTAT data (quarterly national accounts) |
| Job-separation intensity | Percent, Seasonally adjusted | The quarterly transition rate (hazard/Poisson intensity) from employment to unemployment between time $t$ and time $t + 1$ conditional on being employed at time $t$. | Authors' elaboration on ISTAT data (quarterly national accounts) |
| Mature workers' job-finding intensity | Percent, Seasonally adjusted | The transition rate (hazard/Poisson intensity) of workers aged 50 and over from unemployment to employment between time $t$ and time $t + 1$ | Authors' elaboration on ISTAT data (quarterly national accounts) |
| | | conditional on being unemployed at time $t$. | |
| Mature workers' job-separation intensity | Percent, Seasonally adjusted | The transition rate (hazard/Poisson intensity) of workers aged 50 and over from employment to unemployment between time $t$ and time $t+1$ conditional on being employed at time $t$. | Authors' elaboration on ISTAT data (quarterly national accounts) |
| Real government debt | Millions of Chained 2010 Euros, Seasonally Adjusted | Government debt calculated according to the statistical rules established at the European level ("Maastricht debt") | Public finance statistics of Bank of Italy |
| Nominal interest rate | Percent, Seasonally adjusted | Main refinancing operation rate of the European Central Bank (ECB) | Key ECB interest rates statistics released by the ECB |

In Table 2, job-finding intensity and job-separation intensity are calculated from the data on the quarterly transition flows in the labor market provided by ISTAT.

As a first step, job-finding probability and job-separation probability are calculated as below:

$$p_{f,t} = \frac{U_t \rightarrow E_t}{U_t} \tag{63}$$

and:

$$p_{s,t} = \frac{E_t \rightarrow U_t}{E_t} \tag{64}$$

where:

- $p_{f,t}$ is the job-finding probability;
- $p_{s,t}$ is the job-separation probability;
- $U_t \rightarrow E_t$ is the stock of individuals who move from unemployment at time $t-1$ to employment at time $t$;
- $E_t \rightarrow U_t$ is the stock of people who move from employment at time $t-1$ to unemployment at time $t$.

As a second step, job-finding intensity and job-separation intensity are calculated from equations (15) and (16) as below:

$$f_t = -\lg(1 - p_{f,t}) \tag{65}$$

and:

$$s_t = -\lg(1 - p_{s,t}) \tag{66}$$

The deviations of job-separation intensity, job-finding intensity, mature workers' job-finding intensity, mature workers' job-separation intensity, government debt, and nominal interest rate from their respective steady-state values have been calculated by the Hamilton filtering technique (2018), exactly as in the output gap's estimation. The inflation rate in deviation, instead, is calculated as the difference between the actual inflation rate and 0.02 (which represents the average 2% inflation target set by the European Central Bank in the sample period of interest).

Given the heterogeneity in the measurement units, all the variables in Table 2 have been standardized before calculating their statistical moments. This contributes to ensuring the stability of the Differential Evolution algorithm.

## 12. Simulation results

Table 3 reports the observed moments, their counterparts simulated via SMM under Proposition 1, and the difference between the observed and simulated statistics. It is important to remark that the comparisons between the simulated and observed moments are economically meaningful only under the assumption that the structural parameters of the NK model under Proposition 1 lie in the standard domain and do not significantly diverge from those implied by Proposition 3:

**Table 3** Observed and estimated statistical moment under the hypotheses of Proposition 1

| (1) Theoretical moment | (2) Observed moment | (3) Estimated moment | (4) Difference |
|---|---|---|---|
| Mean of real output gap | 0.0000 | 0.1920 | 0.1920 |
| Variance of real output gap | 1.0000 | 0.1100 | -0.8890 |
| Mean of actual inflation rate in deviation | 0.0000 | -0.0704 | -0.0704 |
| Variance of actual inflation rate in deviation | 1.0000 | 0.0539 | -0.9460 |
| Mean job-separation intensity in deviation | 0.0000 | 0.0000 | 0.0000 |
| Variance of job-separation intensity in deviation | 1.0000 | 1.0800 | -0.0753 |
| Mean unemployment rate | 0.0000 | 0.2370 | 0.2370 |
| Variance of unemployment rate | 1.0000 | 0.0261 | -0.9740 |
| Mean real interest rate | 0.0000 | 0.0000 | 0.0000 |
| Variance of real interest rate | 1.0000 | 1.0000 | 0.0000 |
| ACF of real output gap at lag 1 | 0.8410 | 0.9060 | -0.0652 |
| ACF of job-separation intensity in deviation at lag 1 | 0.7350 | 0.9630 | 0.2280 |
| ACF of unemployment rate at lag 1 | 0.9720 | 0.7970 | -0.1740 |
| ACF of real interest rate at lag 1 | 0.2800 | 0.7860 | 0.5060 |
| Correlation between real output gap and job-separation intensity in deviation | 0.6710 | -0.5940 | -1.2600 |
| Correlation between real interest rate and job-separation intensity in deviation | 0.2370 | 0.1690 | -0.0676 |
| Correlation between actual inflation rate in deviation and real output gap | 0.3310 | -0.7740 | -1.1100 |
| Correlation between real interest rate and real output gap | 0.2380 | -0.4320 | -0.6690 |

The simulated output gap displays a higher mean and lower variance than in the data, while both the mean and variance of simulated inflation are below the corresponding observed moments. Within the model, imposing the Taylor principle is associated with a more stable equilibrium path (lower variance) and, in the simulations, with lower inflation variability and a higher average level of aggregate demand relative to the Proposition 3 configuration.

The simulated unemployment rate and job-separation intensity match the observed means but feature lower variances. This pattern is qualitatively consistent with the model's mechanism, whereby labour-market frictions imply that adherence to the Taylor principle (even under full information) reduces unemployment and perceived job-separation risk, while not eliminating either unemployment or separation risk altogether.

Variances and first-order serial correlations in the data differ from those in the simulated series. These discrepancies likely reflect, among other factors, the fact that the reduced-form solution under Proposition 1 differs structurally from that under Proposition 3; in particular, uninformative noise and belief shocks, absent in Proposition 1, can contribute to higher volatility and persistence in Proposition 3.

Simulated cross-correlations are in line with the implications of Proposition 1: under Taylor-rule determinacy, the real interest rate co-moves negatively with inflation and with the output gap. The correlation between job-separation intensity (in deviation) and the output gap is positive, reflecting that, in the model, improved demand/profit conditions are associated with higher hiring and lower separations; conversely, tighter real rates are associated with weaker demand and higher separations. The signs of the observed correlations depart from those reported in column (3) of Table 3, which may reflect the structural differences between Proposition 1 and Proposition 3 and the additional propagation channels present under the latter.

Table 4 reports the results of the same statistics as in Table 3 for the simulation conducted under the hypotheses of Proposition 2 (asymmetric information and Taylor principle):

**Table 4** Observed and estimated statistical moment under the hypotheses of Proposition 2

| (1) Theoretical moment | (2) Observed moment | (3) Estimated moment | (4) Difference |
|---|---|---|---|
| Mean of real output gap | 0.0000 | 0.3280 | 0.3280 |
| Variance of real output gap | 1.0000 | 0.1530 | -0.8470 |
| Mean of actual inflation rate in deviation | 0.0000 | -0.0675 | -0.0675 |
| Variance of actual inflation rate in deviation | 1.0000 | 0.0704 | -0.9300 |
| Mean job-separation intensity in deviation | 0.0000 | 0.0000 | 0.0000 |
| Variance of job-separation intensity in deviation | 1.0000 | 2.0000 | 1.0000 |
| Mean unemployment rate | 0.0000 | 0.0473 | 0.0473 |
| Variance of unemployment rate | 1.0000 | 0.0021 | -0.9980 |
| Mean real interest rate | 0.0000 | 0.0000 | 0.0000 |
| Variance of real interest rate | 1.0000 | 1.0000 | 0.0000 |
| ACF of real output gap at lag 1 | 0.8410 | 0.9060 | 0.0649 |
| ACF of job-separation intensity in deviation at lag 1 | 0.7350 | 0.9670 | 0.2320 |
| ACF of unemployment rate at lag 1 | 0.9720 | 0.7680 | -0.2030 |
| ACF of real interest rate at lag 1 | 0.2800 | 0.8980 | 0.6180 |
| Correlation between real output gap and job-separation intensity in deviation | 0.6710 | -0.5570 | -1.2300 |
| Correlation between real interest rate and job-separation intensity in deviation | 0.2370 | 0.2450 | 0.0086 |
| Correlation between actual inflation rate in deviation and real output gap | 0.3310 | -0.7090 | -1.0400 |
| Correlation between real interest rate and real output gap | 0.2380 | -0.5570 | -0.7950 |

The results reported in Table 4 are qualitatively similar to those in Table 3, but two relevant differences emerge. First, the simulated output gap and actual inflation exhibit higher means and variances relative to their full-information counterparts reported in Table 3. Although the Taylor principle still ensures a unique equilibrium path, asymmetric information implies stronger inflationary pressures and reduced aggregate-output stability compared with the full-information case.

Second, the variance of the simulated job-separation intensity in deviation under Proposition 2 is twice as large as that computed under Proposition 1. The greater dispersion of endogenous macroeconomic variables, most notably actual inflation and the output gap, implied by asymmetric information translates into higher labor-market turnover, as reflected in the increased variance of job-separation intensity, even though the unemployment rate itself remains relatively stable.

In other words, this outcome naturally arises from the private sector's need to form optimal estimates of unobservable fundamentals in an environment characterized by asymmetric information, where the signal provided by public authorities about the current and future state of the economy is contaminated by noise. This additional source of uncertainty hampers firms' ability to efficiently plan production and input choices, including labor demand.

Figure 1 illustrates the job-insecurity measures estimated under Propositions 1 and 2 with second-order risk correction.

**Figure 1** Job insecurity measures simulated under Proposition 1 and Proposition 2

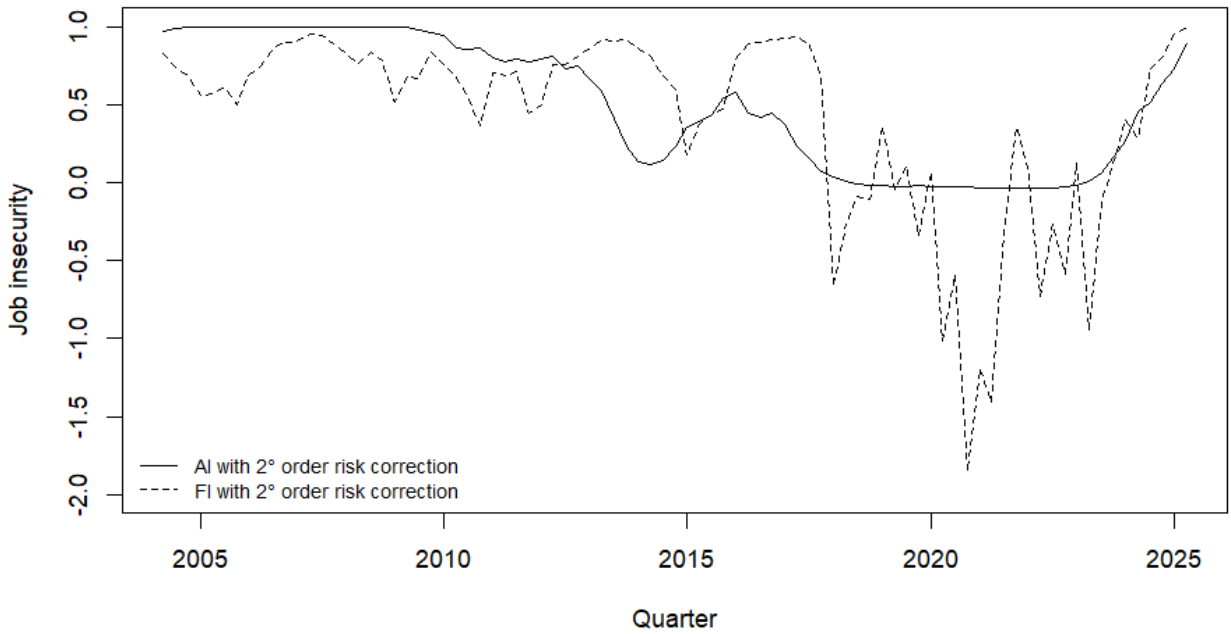

Job insecurity under full information exhibits a lower mean than its asymmetric-information counterpart (0.4093 versus 0.5039), while displaying a higher variance (0.3671 versus 0.1715). The skewness indices of full-information and asymmetric-information job insecurity are equal to -1.6297 and -0.0861, respectively.

Figure 1 illustrates that, under full information, there are several quarters in which the perceived job-separation hazard falls below its long-run benchmark, corresponding to episodes of relatively higher job security. Under asymmetric information, in the model, private agents form optimal estimates via signal extraction (implemented through a Kalman filter in the baseline specification), so that job insecurity appears smoother and only rarely falls below its benchmark. Accordingly, in this specification, asymmetric information is associated with a higher average level of job insecurity.

Figure 2 reports the standardized survey-based index of perceived job insecurity constructed from ISTAT consumer-confidence data. This measure should be interpreted with caution as a proxy for job insecurity under Proposition 3.

**Figure 2** Survey-based index of perceived job insecurity

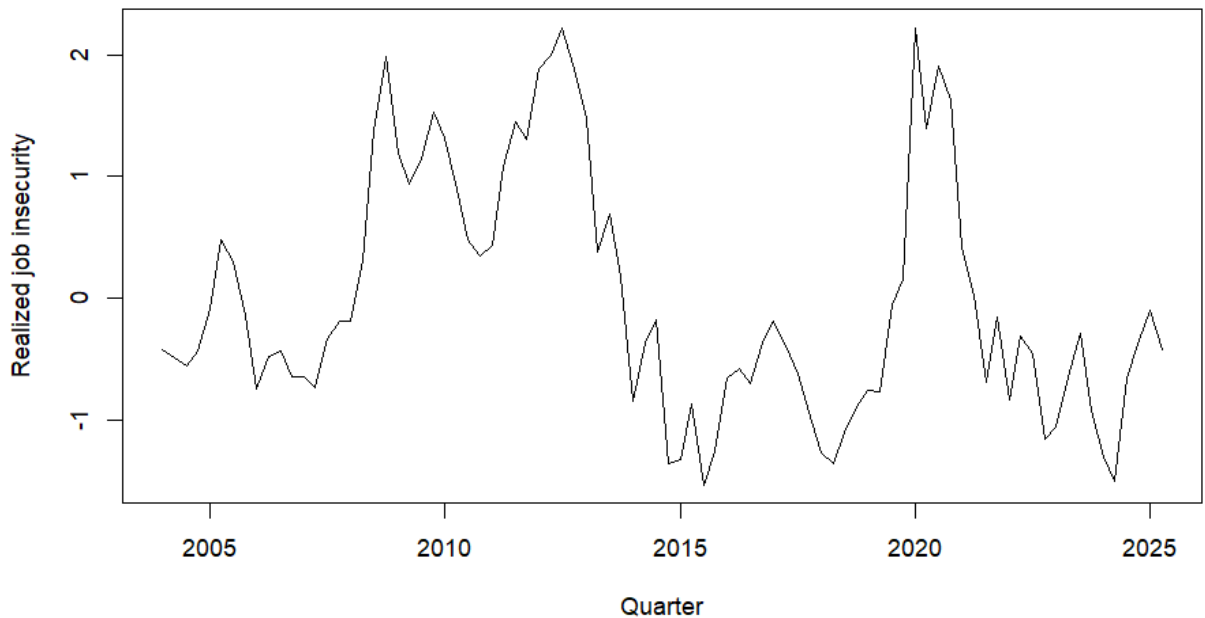

The mean of the unstandardized index is 1.5399, which is dramatically higher than those of the two theoretical cases of Proposition 1 and Proposition 2. Moreover, the skewness of the standardized survey-based index of perceived job insecurity is equal to 0.6608, indicating a right-skewed distribution with relatively frequent moderate values and occasional spikes of very high job insecurity. On average, job insecurity reaches its highest values when the Taylor principle is removed and information is not symmetrically distributed. The comparison between Figure 1 and Figure 2 makes this conclusion visually clear.

Table 5 reports the output of the stochastic volatility tests performed on the three measures of job insecurity via Markov Chain Monte Carlo sampling (Kastner & Frühwirth-Schnatter, 2014; Hosszejni & Kastner, 2019) in which logarithmic volatility is modeled as an AR(1) process with non-zero drift and normally distributed innovations:

**Table 5** Estimated parameters of log-volatility

| (1)<br>Parameter | (2)<br>$JI_t^{[0,12],FI}$ | (3)<br>$JI_t^{[0,12],AI}$ | (4)<br>$\tilde{JI}_t^{[0,12],ISTAT}$ |
|---|---|---|---|
| $\mu_v$ | -0.6130<br>{0.2240} | -1.1200<br>{1.7030} | 4.5600<br>{0.9620} |
| $\varphi_v$ | 0.8280<br>{0.1190} | 0.9700<br>{0.0190} | 0.9000<br>{0.0740} |
| $\sigma_\eta$ | 0.1140<br>{0.0950} | 0.6700<br>{0.1090} | 0.4100<br>{0.1340} |
| $e^{\frac{\mu_v^2}{2}}$ | 0.7410<br>{0.0880} | 0.8900<br>{1.4820} | 10.5100<br>{3.4600} |
| $\sigma_v^2$ | 0.0220<br>{0.0370} | 0.4600<br>{0.1540} | 0.1800<br>{0.1280} |

Standard deviations in curly brackets

The mean of the log stochastic volatility ($\mu_v$) is highest in the asymmetric information, no-Taylor principle scenario, while it is lowest when full information and the Taylor principle hold simultaneously. The degree of persistence of the estimated stochastic volatility ($\varphi_v$), the estimated variance of the stochastic volatility ($\sigma_v^2$), and the standard deviation of the shocks to the estimated stochastic volatility ($\sigma_\eta$) are at their respective maximum values under the hypotheses of Proposition 2, while their minima are recorded under Proposition 1. This result underscores the importance of information hierarchy and transparent institutional communication from the public sector's side to manage and potentially reduce the uncertainty about the job tenure continuity (even when the Taylor principle is imposed) as well as for curbing the adverse impacts of the exogenous shocks on job insecurity. The large values of parameter $\varphi_v$ support the necessity of accounting for the second-order effects when estimating job insecurity.

Table 6 encompasses the estimation output of regression equation (62) with scale parameters:

**Table 6** Estimation output of equation (62)

| (1)<br>Estimated coefficient/test | (2)<br>OLS regression with the "mean-only" job insecurity measures | (3)<br>OLS regression with the 2° order risk correction job insecurity measures |
|---|---|---|
| $\hat{\alpha}_0$ | 0.0000<br>(1779) | 0.0000<br>(0.1775) |
| $\hat{\alpha}_1$ | -0.3721*<br>(0.2165) | -0.3779*<br>(0.2145) |
| $\hat{\alpha}_2$ | 0.5016***<br>(0.1670) | 0.5076***<br>(0.1663) |
| Adjusted $R^2$ | 0.1227 | 0.1256 |
| HAC standard errors | YES | YES |
| Variance Inflation Factor associated with $JI_t^{[0,12],FI}$ | 1.7730 | 1.7820 |
| Variance Inflation Factor associated with $JI_t^{[0,12],AI}$ | 1.7730 | 1.7820 |
| ADF test on $\widetilde{JI}_t^{[0,12],ISTAT}$ | -2.5800<br>[0.1061] | -2.6500<br>[0.0903] |
| ADF test on $JI_t^{[0,12],FI}$ | -2.8857<br>[0.1671] | -2.9050<br>[1.6080] |
| ADF test on $JI_t^{[0,12],AI}$ | 0.1612<br>[0.9978] | 0.0642<br>[0.9970] |
| ADF test on residuals | -2.7400<br>[0.0100] | -2.7300<br>[0.0100] |
| Ramsey RESET test | 0.8247<br>[0.4420] | 0.8064<br>[0.4500] |
| Martingale difference sequence test | 61.0000<br>[0.0980] | 70.3000<br>[0.1100] |

*, **, and *** indicate, respectively, 0.10, 0.05, and 0.01 significance levels
Standard errors in round brackets
p-values in squared brackets

Since the regression model has been estimated on standardized data, the estimated constant $\hat{\alpha}_0$ is zero by construction. The estimated scale parameter $\hat{\alpha}_1$ indicates that the survey-based index of perceived job insecurity is weakly and negatively related to the measure of Proposition 1. On the contrary, the estimated scale parameter $\hat{\alpha}_2$, which is positive and highly significant, indicates that the comovement between the survey-based index of perceived job insecurity and the specific job insecurity measure of Proposition 2 is substantial. This output implies that the Data Generating Process of the actual data (i.e. the job insecurity index constructed using Istat data) closely resembles that of Proposition 2 (asymmetric information and Taylor principle), while being far from that of Proposition 1 (full information with Taylor principle).

The outcomes of the unit-root tests (Augmented Dickey–Fuller tests), conducted on the variables and residuals of regression model (62) using heteroskedasticity- and autocorrelation-consistent (HAC) standard errors, show that the dependent variable (survey-based index of perceived job insecurity) and the independent ones (job insecurity of Proposition 1 and Proposition 2) are cointegrated in the Engle-Granger sense. The output of the Ramsey RESET test justifies the linear specification of the cointegrating equation.

Given the significance levels of estimated coefficients $\hat{\alpha}_1$ and $\hat{\alpha}_2$, one can infer that the long-run cointegrating relationship is mainly led by asymmetric information underlying that DGPs of $JI_t^{[0,12],AI}$ and $\widetilde{JI}_t^{[0,12],ISTAT}$ rather than full information.

The Variance Inflation Factor (VIF) statistics do not point to severe multicollinearity.

The adjusted R-squared confirms that, at least one of the two regressors (presumably $JI_t^{[0,12],AI}$), explains a small but non-negligible part of the variance of $\widetilde{JI}_t^{[0,12],ISTAT}$.

The Chang, Jiang, and Shao (2021) test does not reject the null hypothesis that the regression's residuals follow a martingale difference sequence process, which is consistent with treating $\widetilde{JI}_t^{[0,12],ISTAT}$ as a plausible benchmark of Proposition 3. The insight according to which the DGP of the survey-based index of perceived job insecurity reflects the non-validity of the Taylor principle established by Proposition 3 is broadly consistent with the GMM and 2SLS estimation output of the ECB's Taylor rule (See Appendix A6).

Parameter stability is assessed using the CUSUM test based on the cumulative sum of recursive residuals. The CUSUM path remains within the 95% confidence bands throughout the sample, providing no clear evidence of coefficient instability or discrete structural breaks (See Appendix A7). The results of the same SMM simulation on the mature workers job insecurity index give similar qualitative results (See Section A8 of the final Appendix), confirming the predictive capacity of the NK model extended with cohort heterogeneity, while also delivering some remarkable differences. In particular, the estimated job-separation gap for mature workers displays substantially lower volatility and a much weaker response to macroeconomic shocks, leading to job-insecurity measures that are markedly smoother and closer to their steady-state benchmark compared with the aggregate sample. Moreover, the Kalman-filtered expectations under asymmetric information reveal that older workers update their beliefs more slowly and exhibit a muted sensitivity to informational noise, which results in a significantly smaller discrepancy between the full-information and asymmetric-information job-insecurity estimates. Finally, the regression analysis against realized insecurity shows that the predictive content of the NK model remains intact, but its explanatory power is concentrated in long-run variations rather than short-run fluctuations, indicating that macro-driven job insecurity plays a more structural than cyclical role for this cohort.

## 13. Discussion

The model presented in this work is close in spirit to Han (2024), who analyses the problem of expectation misalignment in a New Keynesian setting in which private agents form projections of unobservable endogenous variables based on a noisy signal about exogenous shocks, while the central bank updates its information set over time through an adaptive learning process. In that study, the issue of equilibrium determinacy is not the main focus, but it is clearly documented that information asymmetry and signal noise can generate economically meaningful forecasting errors among non-professional private agents regarding unobservables. Greater transparency in central-bank communication can therefore help narrow the gap between expected and realised values of the unobservables and, in turn, reduce the likelihood of sizeable deviations from the equilibrium path. The present framework reaches a related qualitative conclusion under a different information hierarchy (with fully informed public agents), while extending the analysis to explicitly consider equilibrium determinacy under alternative communication strategies and the possibility of a "Paradox of Transparency", which is not addressed in Han (2024).

Given its behavioural realism (Evans & Honkapohja, 2009), the adaptive learning approach employed by Han (2024) can be regarded as particularly suitable for capturing aspects of expectation formation that standard rational-expectations models abstract from. At the same time, as emphasised by Eusepi & Preston (2018), learning considerations may constrain the set of policies that can be implemented by central banks, implying that New Keynesian models with adaptive learning need not reproduce all equilibrium outcomes that are feasible under rational expectations. In particular, the complexity of expectation formation under adaptive learning may be associated with local instability or with equilibria that are admissible in principle but not aligned with the central bank's target. In light of this concern, the analysis complements the rational-expectations results with a robustness check under adaptive learning with decreasing gain.

Within the rational-expectations New Keynesian model developed here (with fully informed public agents and partially informed households and firms) adherence to the Taylor principle emerges as both a necessary and sufficient condition for equilibrium determinacy. This implication differs from Sorge & Vota (2025), who study equilibrium characterisation under an alternative information hierarchy motivated by Lubik et al. (2023). In the baseline IS–NKPC–Taylor-rule framework considered by Sorge & Vota (2025), asymmetric information can generically give rise to multiple linear sunspot equilibria even when the Taylor principle holds, so that the Taylor principle is no longer sufficient for determinacy. The divergence appears to reflect differences in model structure: Sorge & Vota (2025) operate within a three-equation New Keynesian framework featuring monetary-policy opacity (the private sector does not directly observe the policy rate and forms projections), whereas the present paper augments the baseline block with labour-market frictions, fiscal policy, and a larger set of persistent exogenous and belief shocks. In this richer environment, multiple sunspot belief shocks arise only when the Taylor principle is relaxed; under either full or asymmetric information combined with Taylor-rule determinacy, the equilibrium path remains unique.

Regarding the "Paradox of Transparency", Sánchez (2013) shows that the phenomenon can result from the disclosure of the central bank's preferences over inflation stabilisation and economic growth and from the resulting strategic interaction between the central bank and the private sector. The present framework suggests that, even when the central bank's preferences (as summarised by the policy coefficients of the Taylor rule) are perfectly known to households and firms and the informative signal is noise-free, transparency-related paradoxical effects may still arise. In the model, this outcome reflects the interaction between news shocks and other innovations rather than strategic preference revelation.

Taken together, the theoretical analysis indicates that an exclusively individual-based approach may not be sufficient to fully assess the magnitude and determinants of job insecurity when the latent component of perceived dismissal risk is influenced by persistent macroeconomic disturbances, rather than solely by unpredictable idiosyncratic shocks. Similarly, non-structural empirical specifications aiming to quantify the marginal effects of macroeconomic variables on job insecurity (Ellonen & Nätti, 2015; Johnston et al., 2020) may be vulnerable to bias or misspecification if they abstract from the persistence of exogenous shocks (including news shocks), forward-looking optimal behaviour, the monetary-policy regime, and the possibility of belief-driven components that may arise when the Taylor principle is violated, as well as from regime-dependent volatility in job insecurity. In this sense, structural restrictions and equilibrium characterisation provide a useful discipline for interpreting job-insecurity dynamics.

Several contributions stress the role of transparency in reducing inflation and stabilising inflation expectations, largely through credibility gains associated with clear communication and the revelation of preferences over output and price stability, while others argue that some degree of opaqueness may be preferable for achieving price-stability objectives (Weber, 2016). The present manuscript highlights that, within the proposed framework, equilibrium determinacy is primarily tied to adherence to the Taylor principle, without implying that greater transparency of the informative signal necessarily entails welfare gains for households and firms.

The empirical patterns and simulation results reported in this paper indicate that the interaction between informational frictions, Taylor-type monetary policy, and (when admissible) sunspot belief shocks can generate regime-dependent levels and dynamics of job insecurity. Under full information and Taylor-rule determinacy, job-separation risk remains relatively contained and closely linked to fundamentals; under asymmetric information and a deviation from the Taylor principle, the combination of noisy signals and extrinsic belief shocks can produce higher and more volatile job insecurity, in a way that is qualitatively consistent with salient features of the ISTAT-based survey proxy.

A growing body of empirical and survey evidence documents that individuals tend to overestimate subjectively perceived job-separation intensity when facing noisy and ambiguous information about the state of the economy and elevated uncertainty (Dominitz & Manski, 1997; Hendren, 2017; Mueller & Spinnewijn, 2023; Roth & Wohlfart, 2020; Faberman et al., 2022). The structural NK model developed here links job insecurity to equilibrium properties under alternative monetary-policy regimes and information hierarchies (full information versus different degrees of asymmetric information). While several macroeconomic models incorporate incomplete information and signal extraction (Lorenzoni, 2009; Angeletos & Lian, 2018), the present analysis contributes by connecting monetary-policy information frictions to survey-based measures of perceived job-separation risk.

Within the model, asymmetric information implies that excess sensitivity in job insecurity reflects both fundamental and non-fundamental components. The latter take the form of belief shocks that may arise when the Taylor principle is violated and are associated with the filtration of noise contaminating the informative signal. In this environment, systemic misperception can emerge endogenously as a distortion in the estimated job-separation process induced by signal noise. Moreover, the mean and volatility of the underlying process affect the transmission of innovations to job-separation, generating non-linearities in the mapping from shocks to perceived dismissal risk.

A further implication is that, in the presence of noisy signals and news shocks, the timing of the policy-rate announcement is not quantitatively central for job-insecurity dynamics in the baseline specification. This result is conditional on the maintained assumption that news about current fundamentals can be distinguished from news about future fundamentals, which permits agents to disentangle informative components from uninformative noise and to form projections for nominal (and therefore real) interest rates. Under these assumptions, projections can be constructed even with a parsimonious information set consisting of the history of the informative signal.

The cohort-specific analysis suggests that mature workers (aged 50 and above) experience lower and less volatile job insecurity than the aggregate labour force, consistent with evidence that job-separation hazards, occupational mobility, and contract turnover decline over the life cycle. Within the structure of the model, this pattern is accounted for by two mechanisms: i) relatively smaller deviations in job-finding and job-separation rates for mature workers, and *ii*) higher persistence but lower dispersion of the fundamental shocks affecting their labour-market transitions. As a result, expectations for this cohort respond more weakly to informational noise and macroeconomic innovations, producing a muted transmission of asymmetric-information shocks to job insecurity.

Even though the level and volatility of job insecurity are lower for this group, the asymmetric-information equilibrium still amplifies perceived dismissal risk relative to the full-information case. This pattern is consistent with the view that informational frictions can represent a structural source of perceived uncertainty in labour markets even among cohorts typically characterised by lower turnover. Consequently, the model provides a coherent rationale for why older workers may remain less sensitive to aggregate fluctuations while still being exposed, albeit to a milder extent, to signal-extraction distortions in forming expectations about job stability.

## 14. Concluding remarks and policy implications

Departing from the conventional approach based on individual and meso variables, this manuscript has extended the analysis of job insecurity to its systemic dimension, in particular the macroeconomic one. To accomplish this research goal, the paper has developed a New Keynesian model with asymmetric information and a noisy informative signal in which the standard IS-NKPC-Taylor rule block is enriched with a modern search-and-matching framework for the labour market.

Job insecurity has been characterized across different equilibria, each of which results from a specific combination of information hierarchy and monetary policy regime.

The theoretical and empirical results have demonstrated that the response of job insecurity to a certain exogenous shock is always nonlinear and often determined by the complex interaction with other disturbances. Moreover, job insecurity has been found to display stochastic volatility, even when it is hit by homoscedastic shocks. It has also been demonstrated that, under asymmetric information and with no aggressive monetary policy, the (eventual) equilibrium path of job insecurity may not be sunspot-free.

The natural policy recommendation emerging from the present manuscript is that reducing the noisy content of the informative signal (namely, adopting a transparent institutional communication) is crucial to dampen and stabilize job insecurity around its long-run, steady-state benchmark. However, decision makers need to carefully ponder the opportunity of releasing unexpected information on the current or future state of the economy (news shocks). In fact, because of the tension between the demand and central bank policy reaction effects, positive informative shocks can boost the expected job-separation intensity instead of reducing it.

The findings related to the mature workers group allow one to trace out some qualitative indications for this specific fragile category (which, however, require further analysis for rigorous confirmation). First, since asymmetric information raises job insecurity among mature workers despite their inherently low separation risk, clearer forward guidance and improved communication about the state of the economy may reduce unnecessary precautionary reactions near retirement. Second, because the baseline separation hazard for mature workers is already low and weakly cyclical, strengthening the employment protection legislation for this group produces limited gains in perceived security. Policies aimed at information transparency are likely more effective. Third, given that mature workers update expectations mainly through macro signals (not through realized transitions), the information-intensive active labour market policies (such as skill assessments, job-transition counselling, early-warning systems) may yield higher welfare gains than traditional hiring subsidies. Fourth, due to the fact that mature workers form expectations over longer horizons with low volatility, adjustments in retirement age or pension rules can shift perceived job stability through expectations rather than through transition risk. Policymakers should therefore account for expectation channels when evaluating reforms targeting older workers.

Thus, central banks are called upon to conduct an ex-ante evaluation of the demand channel of the news shocks and the related central bank's feedback response before revealing to the public unexpected information on the fundamentals.

A promising direction for the subsequent contributions would be augmenting the NK model proposed in this paper with unemployment risk aversion. Another valuable avenue for future research would be incorporating sectoral heterogeneity in the model to uncover the similarities and differences in terms of job insecurity between industries (a relevant point in the literature on this subject matter.

# Appendix

## *A1. Proof of Proposition 1 and Proposition 2*

The state-space representation of the model is given by:

$$\Phi_0 E_t\left[\gamma_{t+1}^{\Re} | \mathcal{Q}_t\right] = \Phi_1 \gamma_t + \Phi_2 \Xi_t$$

where:

$$\gamma_t = \begin{bmatrix} \hat{\pi}_{t+1} \\ \hat{y}_{t+1} \end{bmatrix}, \qquad \Phi_0 = \begin{bmatrix} \beta & 0 \\ -\dfrac{1}{\sigma} & -1 \end{bmatrix}, \qquad \Phi_1 = \begin{bmatrix} 1 & k \\ \dfrac{\alpha_\pi}{\sigma} & \dfrac{\alpha_y + \sigma}{\sigma} \end{bmatrix}, \qquad \Phi_2 = \begin{bmatrix} 1 & 0 & 0 & 0 & 0 \\ 0 & \dfrac{1}{\sigma} & -\dfrac{1}{\sigma} & -\delta_a & -\delta_G \end{bmatrix}, \qquad \Xi_t = \begin{bmatrix} \varepsilon_t^\pi \\ \varepsilon_t^i \\ \bar{r}_t \\ a_t \\ \hat{g}_t \end{bmatrix}$$

In order to determine the number of solutions of the structural form of the model, one needs to count the number of stable eigenvalues of transition matrix $A = \Phi_0^{-1}\Phi_1$ and compare it with the number of forward-looking variables of the state-space representation (two). If the two numbers are equal, then the model admits equilibrium determinacy and uniqueness.

Under the condition $\sigma + \alpha_y - \alpha_\pi k \neq 0$, which grants $\det(\Phi_0) \neq 0$, the eigenvalues of $A$ are equal to:

$$\Omega_{1,2} = \frac{\sigma + \beta(\sigma + \alpha_y) - k \pm \sqrt{[\sigma + \beta(\sigma + \alpha_y) - k]^2 - 4\sigma\beta(\sigma + \alpha_y - \alpha_\pi k)}}{2(\sigma + \alpha_y - \alpha_\pi k)}$$

If $\Delta = [\sigma + \beta(\sigma + \alpha_y) - k]^2 - 4\sigma\beta(\sigma + \alpha_y - \alpha_\pi k) < 0$, the two eigenvalues are complex conjugates whose absolute values are given by $|\Omega_{1,2}| = \sqrt{\frac{\sigma\beta}{\sigma + \alpha_y - \alpha_\pi k}}$.

If $\Delta \geq 0$, instead, the two eigenvalues belong to the real number set ($\mathbb{R}$), and their absolute values are equal to $|\Omega_{1,2}| = \left| \frac{\sigma + \beta(\sigma + \alpha_y) - k \pm \sqrt{[\sigma + \beta(\sigma + \alpha_y) - k]^2 - 4\sigma\beta(\sigma + \alpha_y - \alpha_\pi k)}}{2(\sigma + \alpha_y - \alpha_\pi k)} \right|$.

Upon setting $\alpha_y \geq 0$, $\sigma > 0$, $k > 0$, $0 < \beta < 1$ and $\alpha_\pi > 1$, both $\Omega_1$ and $\Omega_2$ fall inside the unit circle; namely, the two eigenvalues are stable. This implies that the Blanchard-Khan (1980) condition for equilibrium determinacy and uniqueness is satisfied.

The proof of Proposition 2 follows the same steps, as, after switching off the exogenous shifters, its state-space representation coincides with that of the full information case.

### A2. Proof of approximation (34)

Upon defining the integral of job-separation intensity at time $t + h$ with $h = H\Delta$ as $\mathrm{I}_t = -\int_t^{t+H\Delta} \hat{s}_\tau^{\mathfrak{R}} d_\tau$, equation (33) can be rearranged as:

$$JI_t^{[0,h],\mathfrak{R}} = 1 - E_t[e^{\mathrm{I}_t}|\mathcal{Q}_t]$$

Under the assumption that the fluctuations of $\mathrm{I}_t$ around its conditional mean $\mu_{I,t} = E_t[\mathrm{I}_t|\mathcal{Q}_t]$ are negligible (small volatility), it is possible to express $I_t$ as:

$$I_t = \mu_{I,t} + \tilde{I}_t$$

with $E_t[\tilde{I}_t|\mathcal{Q}_t] = 0$ and then one can calculate the Taylor expansion of $e^{\mathrm{I}_t}$ around $\mu_{I,t}$:

$$e^{\mathrm{I}_t} = e^{\mu_{I,t} + \tilde{I}_t} = e^{\mu_{I,t}} e^{\tilde{I}_t} \approx e^{\mu_{I,t}} \left( 1 + \tilde{I}_t + \frac{1}{2}\tilde{I}_t^2 + \cdots \right)$$

from which:

$$E_t[e^{\mathrm{I}_t}|\mathcal{Q}_t] \approx e^{\mu_{I,t}} = e^{E_t[I_t|\mathcal{Q}_t]} \tag{a}$$

By replacing the approximation above in equation (33), one obtains:

$$JI_t^{[0,h],\mathfrak{R}} \approx 1 - e^{E_t[I_t|\mathcal{Q}_t]} = 1 - e^{-E_t\left[\int_t^{t+H\Delta} \hat{s}_\tau^{\mathfrak{R}} d_\tau | \mathcal{Q}_t\right]}$$

In addition, by using the Riemann sum, one can make the following approximation:

$$E_t\left[ \int_t^{t+H\Delta} \hat{s}_\tau^{\mathfrak{R}} d_\tau | \mathcal{Q}_t \right] \approx \Delta \sum_{h=1}^H E_t[\hat{s}_{t+h\Delta}^{\mathfrak{R}}|\mathcal{Q}_t] = \Lambda_t(h)^{\mathfrak{R}} \tag{b}$$

Thus, $JI_t^{[0,h],\mathfrak{R}}$ becomes:

$$JI_t^{[0,h],\Re} \approx 1 - e^{-A_t(h)^{\Re}}$$

which is exactly approximation (34).

The total approximation error in equation (34) is generated by (i) error $\varepsilon_t^{dis}$ due to the use of the discretization of the integral in equation (a), namely, $E_t[I_t|Q_t] - E_t[\tilde{I}_t|Q_t]$, and (ii) the nonlinearity of the exponential function when replacing $I_t$ with $\tilde{I}_t$ as in equation (b) for the purpose of calculating $JI_t^{[0,h],\Re}$ in equation (34):

$$\varepsilon_t^{nonlin} = [1 - e^{-I_t}] - [1 - e^{-\tilde{I}_t}] = e^{-c_t}(I_t - \tilde{I}_t)$$

The formula above indicates that $\varepsilon_t^{nonlin}$ is proportional to the error on the integral for some $c_t$ such that $e^{-c_t} \in (0,1]$.

Hence, one needs to prove that the size orders of both discretization error $\varepsilon_t^{dis}$ and nonlinearity error $\varepsilon_t^{nonlin}$ are negligible at least for a plausible range of values of $\hat{s}_\tau^{\Re}$, $H$, and $\Delta$.

If $\hat{s}_\tau^{\Re}$ is regular in $\tau$ (which is consistent with the assumption that it is the projection of a stationary ARX(1) process), $\varepsilon_t^{dis}$ is subjected to the standard bound:

$$\left|\varepsilon_t^{dis}\right| \leq \frac{H\Delta}{2} \sup_{\tau \in [t, t+H]} \left|\hat{s}_{\tau|t}^{\Re}{}'\right|$$

where $\hat{s}_{\tau|t}^{\Re}{}'$ is the speed at which job-separation intensity in deviation varies over time.

For $\Delta = \frac{1}{4}$ (quarterly data), $H = 8$ (two years), $\sup_{\tau \in [t, t+H]} \left|\hat{s}_{\tau|t}^{\Re}{}'\right| = 0.02$ (very gradual changes compatible with the modelling of $\hat{s}_\tau^{\Re}$ as a stationary autoregressive process), and $\rho_s = 0.80$, one obtains $\left|\varepsilon_t^{dis}\right| \leq 10^{-2}$.

Error $\left|\varepsilon_t^{dis}\right|$ transmits to $JI_t^{[0,h],\Re}$ according to the following equation:

$$\left|\varepsilon_t^{nonlin}\right| = \left|e^{-c_t}\right|\left|\varepsilon_t^{dis}\right| \leq \left|\varepsilon_t^{dis}\right|$$

Then, for the same set of values as those used in the calculation of $\left|\varepsilon_t^{dis}\right|$, the bound on $\left|\varepsilon_t^{nonlin}\right|$ is even lower than 0.01.

### A3. Proof of equation (36)

Suppose that the shocks of the linearized version of the New Keynesian model are normally distributed and define $\hat{S}_{N,t}^{\Re} = -\int_t^{t+h} \hat{s}_\tau^{\Re} d_\tau$ with $\hat{S}_{N,t}^{\Re}|\Theta_t \sim N(\mu_{S,t}^{\Re}, \sigma_{S,t}^{2,\Re})$.

Thus, rearrange equation (33) as

$$JI_t^{[0,h],\Re} = 1 - E_t\left[e^{\hat{S}_{N,t}^{\Re}}|Q_t\right]$$

By using the moment generating function of the normal distribution, one can calculate $E_t\left[e^{k\hat{S}_{N,t}^{\Re}}|Q_t\right]$ as:

$$E_t\left[e^{k\hat{S}_{N,t}^{\Re}}|Q_t\right] = e^{\left(k\mu_{S,t}^{\Re} + \frac{1}{2}k^2\sigma_{S,t}^{2,\Re}\right)}$$

Where $k \in \mathbb{R}$.

For $k = 1$, one obtains:

$$E_t\left[e^{k\hat{S}_{N,t}^{\Re}}|Q_t\right] = e^{\left(-\mu_{S,t}^{\Re} + \frac{1}{2}\sigma_{S,t}^{2,\Re}\right)}$$

Finally, substituting out the previous expression in equation (33) yields:

$$JI_t^{[0,h],\mathfrak{R}} = 1 - e^{\left(-\mu_{S,t}^{\mathfrak{R}} + \frac{1}{2}\sigma_{S,t}^{2,\mathfrak{R}}\right)}$$

which is exactly equation (36).

### A4. Proof of Proposition 3

Given the innovation of the informative signal (its exogenous, unpredictable part which is unexplained by the fundamentals)

$$\Gamma_t = a_t - E_{t-1}[a_t|\mathcal{H}_{t-1}]$$

assume that the private agents' expectations are anchored to an extrinsic driver $\mathbb{Z}_t$ (sunspot) which admits the following AR(1) representation:

$$\mathbb{Z}_t = \rho_{\mathbb{Z}}\mathbb{Z}_{t-1} + \Gamma_t$$

with $|\rho_{\mathbb{Z}}| < 1$.

Si ipotizzi altresì che il modello ammetta una soluzione del tipo: Furthermore, assume that the model admits a solution of the type:

$$\hat{y}_t = b_1\mathbb{Z}_t$$

and:

$$\hat{\pi}_t = b_2\mathbb{Z}_t$$

from which:

$$E_t[\hat{y}_{t+1}|\mathcal{H}_t] = b_1\rho_{\mathbb{Z}}\mathbb{Z}_t$$

and:

$$E_t[\hat{\pi}_{t+1}|\mathcal{H}_t] = b_2\rho_{\mathbb{Z}}\mathbb{Z}_t$$

Thus, the state-space representation of the model with asymmetric information becomes:

$$\begin{cases} \left(1 + \frac{\alpha_y}{\sigma} - \rho_{\mathbb{Z}}\right)b_1 + \left(-\frac{\alpha_\pi}{\sigma} + \frac{\rho_{\mathbb{Z}}}{\sigma}\right)b_2 = 0 \\ -kb_1 + (1 - \beta\rho_{\mathbb{Z}}) = 0 \end{cases}$$

The existence of a non-trivial solution to the system above $((b_1, b_2) \neq (0,0))$ requires the satisfaction of the compatibility condition:

$$C(\rho_{\mathbb{Z}}) = \left(1 + \frac{\alpha_y}{\sigma} - \rho_{\mathbb{Z}}\right)(1 - \beta\rho_{\mathbb{Z}}) - k\left(\frac{\alpha_\pi}{\sigma} - \frac{\rho_{\mathbb{Z}}}{\sigma}\right) = 0$$

$C(\rho_{\mathbb{Z}})$ is evaluated in the points $\rho_{\mathbb{Z}} = 0$:

$$C(0) = 1 + \frac{\alpha_y}{\sigma} - \frac{k\alpha_\pi}{\sigma}$$

and $\rho_{\mathbb{Z}} = \alpha_\pi$:

$$C(\alpha_\pi) = \left(1 + \frac{\alpha_y}{\sigma} - \alpha_\pi\right)(1 - \beta\alpha_\pi) \geq 0$$

If $\alpha_\pi > \frac{\sigma+\alpha_y}{k}$ and $\alpha_\pi \leq 1$, then $C(0) < 0$, and, by continuity (zero theorem), there exists at least one value of $\rho_{\mathbb{Z}} \in (0, \alpha_\pi) \subset (-1, +1) | C(\rho_{\mathbb{Z}}) = 0$.

### A5. Proof of Proposition 4

Define the latent variables relevant for the information signal as:

$$\mathbb{X}_t = \begin{bmatrix} q_t \\ \varepsilon_t^b \end{bmatrix}$$

where, as specified in Section 3:

$$q_t = \rho_q q_{t-1} + \lambda_t, \qquad |\rho_q| < 1, \qquad \lambda_t \sim i.i.d. \left(0, \sigma_\lambda^2\right)$$

$$\varepsilon_t^b = \rho_b \varepsilon_{t-1}^b + \xi_t, \qquad |\rho_b| < 1, \qquad \xi_t \sim i.i.d. \left(0, \sigma_\xi^2\right)$$

and the overall informative signal is given by:

$$a_t = q_t + \varepsilon_t^b + v_t, \qquad v_t \sim i.i.d. \left(0, \sigma_v^2\right)$$

The state-space representation for $\mathbb{X}_t$ is given by:

$$\mathbb{X}_t = A_{\mathbb{X}} \mathbb{X}_{t-1} + \mathbb{u}_t, \qquad A_{\mathbb{X}} = \begin{bmatrix} \rho_q & 0 \\ 0 & \rho_b \end{bmatrix}, \qquad \mathbb{u}_t = \begin{bmatrix} \lambda_t \\ \xi_t \end{bmatrix}$$

$$a_t = \mathbb{C}_{\mathbb{X}} \mathbb{X}_t + v_t, \qquad \mathbb{C}_{\mathbb{X}} = \begin{bmatrix} 1 & 1 \end{bmatrix}$$

For $\rho_q \neq \rho_b$, the pair $(A_{\mathbb{X}}, \mathbb{C}_{\mathbb{X}})$ is observable and the observability matrix is as below:

$$O_{\mathbb{X}} = \begin{bmatrix} \mathbb{C}_{\mathbb{X}} \\ \mathbb{C}_{\mathbb{X}} A_{\mathbb{X}} \end{bmatrix} = \begin{bmatrix} 1 & 1 \\ \rho_q & \rho_b \end{bmatrix}$$

and it has full rank 2. Thus, there exists a unique linear minimum mean square error estimator of $\mathbb{X}_t$ based on the history $\{a_\tau\}_{\tau \leq t}$, and private agents can implement a Kalman filter to recover optimal estimates of both $q_t$ and $\varepsilon_t^b$ from the signal $a_t$.

In this case, the minimum information set for Kalman filtering of the fundamentals is given by

$$\mathcal{H}_t^{\min} = \hbar(\{a_\tau\}_{\tau \leq t})$$

i.e. the history of the composite signal alone is informationally sufficient to render the equilibrium in Proposition 2 well defined under asymmetric information.

### A6 GMM and 2SLS estimates of the ECB's Taylor rule

**Table A** GMM and TSLS estimates of the ECB's Taylor rule (2004Q1 − 2025Q2)

| Coefficient/Test | GMM | 2SLS |
|---|---|---|
| Const | 0.0914 | 0.0000 |
|  | (0.4065) | (0.0906) |
|  |  |  |
| Inflation rate | 0.0773 | 0.1382 |
|  | (0.1204) | (0.1072) |
| Output gap | 0.4422 | 0.3977*** |
|  | (0.2843) | (0.1072) |
| Sargan test | 0.7957 | 3.2668 |
|  | [0.3675] | [0.0706] |

*, **, and *** indicate, respectively, 0.10, 0.05, and 0.01 significance levels
Standard errors in round brackets
p-values in squared brackets
Instrument set: Inflation rate at lag 1, output gap at lag 1, and Brent crude-oil price

## A7. CUSUM charts of estimated regression equation (62)

**Figure A** CUSUM chart for regression equation (62) estimated on the "mean only" job insecurity measures

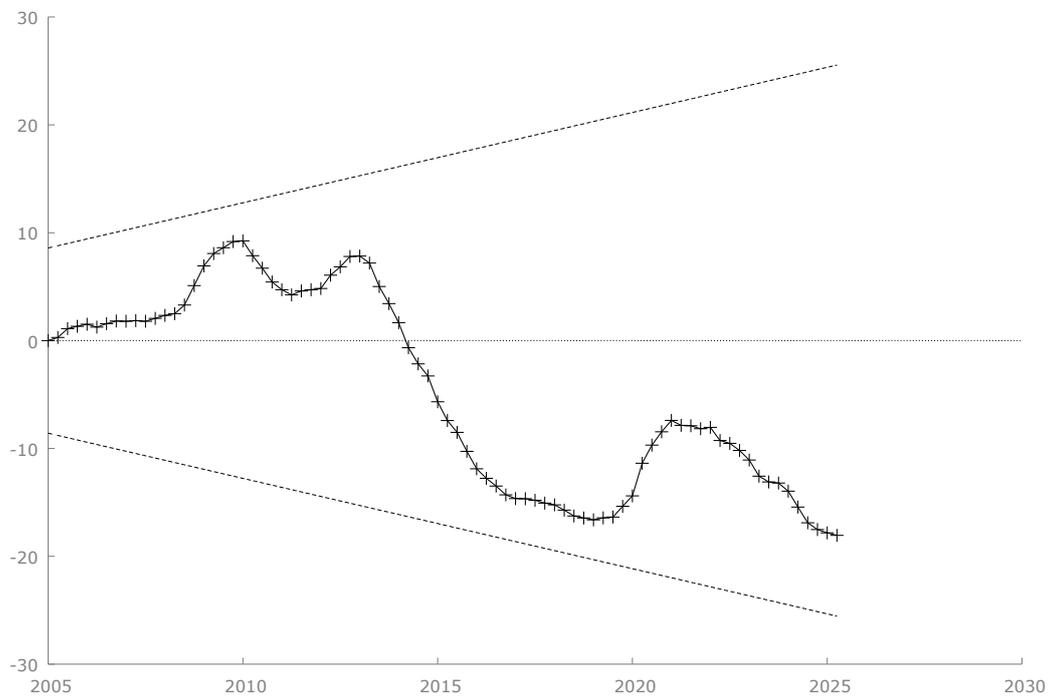

**Figure B** CUSUM chart for regression equation (62) estimated on the job insecurity measures incorporating the 2° order risk correction

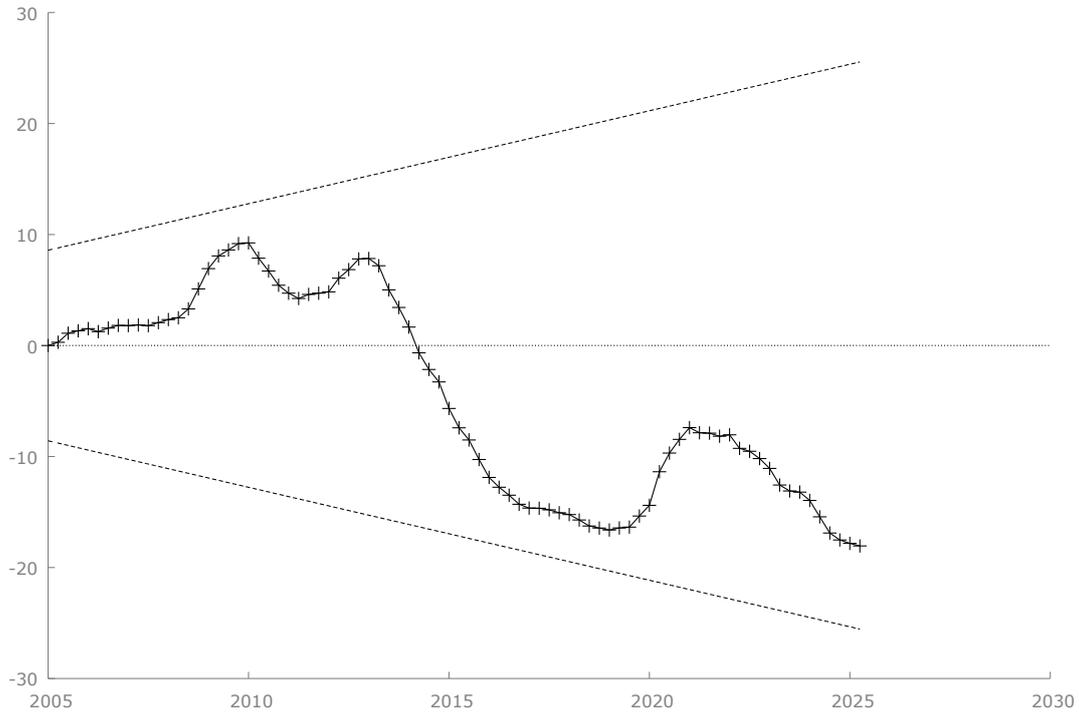

## A8. SMM simulation of the model's version with cohort heterogeneity

**Table B** Coefficients used for calibration purposes under Proposition 1 and Proposition 2 with cohort heterogeneity

| (1) Parameter | (2) Value under Proposition 1 with mature workers | (3) Value under Proposition 2 with mature workers |
|---|---|---|
| $\beta$ | 0.99 | 0.99 |
| $k$ | 0.10 | 0.10 |
| $\sigma$ | 1.00 | 1.00 |
| $\alpha_y$ | 0.10 | 0.10 |
| $\alpha_\pi$ | 1.15 | 1.15 |
| $\phi_y$ | 0.00 | 0.00 |
| $\phi_r$ | 0.00 | 0.00 |
| $\psi_y$ | 0.20 | 0.20 |
| $\psi_r$ | 0.20 | 0.20 |
| $\delta_g$ | 0.25 | 0.25 |
| $\delta_\lambda$ | 0.35 | 0.35 |
| $\delta_\xi$ | 0.35 | 0.35 |
| $\rho_\epsilon$ | 0.63 | 0.58 |
| $\rho_g$ | 0.29 | 0.12 |
| $\rho_i$ | 0.29 | 0.57 |
| $\rho_{\hat{y}}$ | 0.23 | 0.35 |
| $\rho_{\bar{y}}$ | 0.18 | 0.43 |
| $\rho_b$ | 0.93 | 0.94 |
| $\rho_q$ | 0.93 | 0.91 |
| $\rho_f$ | 0.24 | 0.42 |
| $\rho_s$ | 0.29 | 0.92 |
| $\sigma_\lambda^2$ | 0.10 | 0.10 |
| $\sigma_\xi^2$ | 0.10 | 0.10 |
| $\sigma_\omega^2$ | 0.20 | 0.20 |
| $\sigma_\vartheta^2$ | 0.30 | 0.30 |
| $\sigma_\zeta^2$ | 0.20 | 0.20 |
| $\sigma_w^2$ | 0.10 | 0.10 |
| $\sigma_f^2$ | 0.05 | 0.05 |
| $\sigma_s^2$ | 0.05 | 0.05 |
| $\sigma_\eta^2$ | 0.20 | 0.20 |

**Table C** Observed and estimated statistical moment under the hypotheses of Proposition 1 with cohort heterogeneity

| (1)<br>*Theoretical moment* | (2)<br>*Observed moment* | (3)<br>*Estimated moment* | (4)<br>*Difference* |
|---|---|---|---|
| Mean of real output gap | 0.0000 | 0.4360 | 0.4360 |
| Variance of real output gap | 1.0000 | 0.5360 | 0.5360 |
| Mean of actual inflation rate in deviation | 0.0000 | -0.2090 | -0.2090 |
| Variance of actual inflation rate in deviation | 1.0000 | 0.2470 | 0.2470 |
| Mean job-separation intensity in deviation | 0.0000 | 0.0000 | 0.0000 |
| Variance of job-separation intensity in deviation | 1.0000 | 0.9790 | 0.0123 |
| Mean unemployment rate | 0.0000 | 0.2840 | 0.2840 |
| Variance of unemployment rate | 1.0000 | 0.0258 | -0.9740 |
| Mean real interest rate | 0.0000 | 0.0000 | 0.0000 |
| Variance of real interest rate | 1.0000 | 1.0000 | 0.0000 |
| ACF of real output gap at lag 1 | 0.8410 | 0.9570 | 0.1160 |
| ACF of job-separation intensity in deviation at lag 1 | 0.3400 | 0.9850 | 0.6450 |
| ACF of unemployment rate at lag 1 | 0.9030 | 0.8550 | -0.0477 |
| ACF of real interest rate at lag 1 | 0.2800 | 0.5370 | 0.2570 |
| Correlation between real output gap and job-separation intensity in deviation | -0.4110 | -0.5120 | -0.1010 |
| Correlation between real interest rate and job-separation intensity in deviation | -0.1210 | -0.2540 | -0.1340 |
| Correlation between actual inflation rate in deviation and real output gap | 0.3310 | -0.8680 | -1.200 |
| Correlation between real interest rate and real output gap | 0.2380 | -0.0706 | -0.3080 |

**Table D** Observed and estimated statistical moment under the hypotheses of Proposition 1 with cohort heterogeneity

| (1)<br>*Theoretical moment* | (2)<br>*Observed moment* | (3)<br>*Estimated moment* | (4)<br>*Difference* |
|---|---|---|---|
| Mean of real output gap | 0.0000 | 0.3940 | 0.3940 |
| Variance of real output gap | 1.0000 | 0.4510 | -0.5490 |
| Mean of actual inflation rate in deviation | 0.0000 | -0.1780 | -0.1780 |
| Variance of actual inflation rate in deviation | 1.0000 | 0.2030 | -0.7970 |
| Mean job-separation intensity in deviation | 0.0000 | 0.0000 | 0.0000 |
| Variance of job-separation intensity in deviation | 1.0000 | 0.9890 | -0.0112 |
| Mean unemployment rate | 0.0000 | 0.2660 | 0.2660 |
| Variance of unemployment rate | 1.0000 | 0.0232 | -0.9770 |
| Mean real interest rate | 0.0000 | 0.0000 | 0.0000 |
| Variance of real interest rate | 1.0000 | 1.0000 | 0.0000 |
| ACF of real output gap at lag 1 | 0.8410 | 0.9500 | 0.1100 |
| ACF of job-separation intensity in deviation at lag 1 | 0.3400 | 0.9830 | 0.6440 |
| ACF of unemployment rate at lag 1 | 0.9030 | 0.9150 | 0.0120 |
| ACF of real interest rate at lag 1 | 0.2800 | 0.7130 | 0.4330 |
| Correlation between real output gap and job-separation intensity in deviation | -0.4110 | -0.4250 | -0.0135 |
| Correlation between real interest rate and job-separation intensity in deviation | -0.1210 | -0.1860 | -0.0655 |
| Correlation between actual inflation rate in deviation and real output gap | 0.3310 | -0.8500 | -1.1800 |
| Correlation between real interest rate and real output gap | 0.2380 | -0.1620 | -0.3990 |

**Figure E** Job insecurity of mature workers simulated under Proposition 1 and Proposition 2

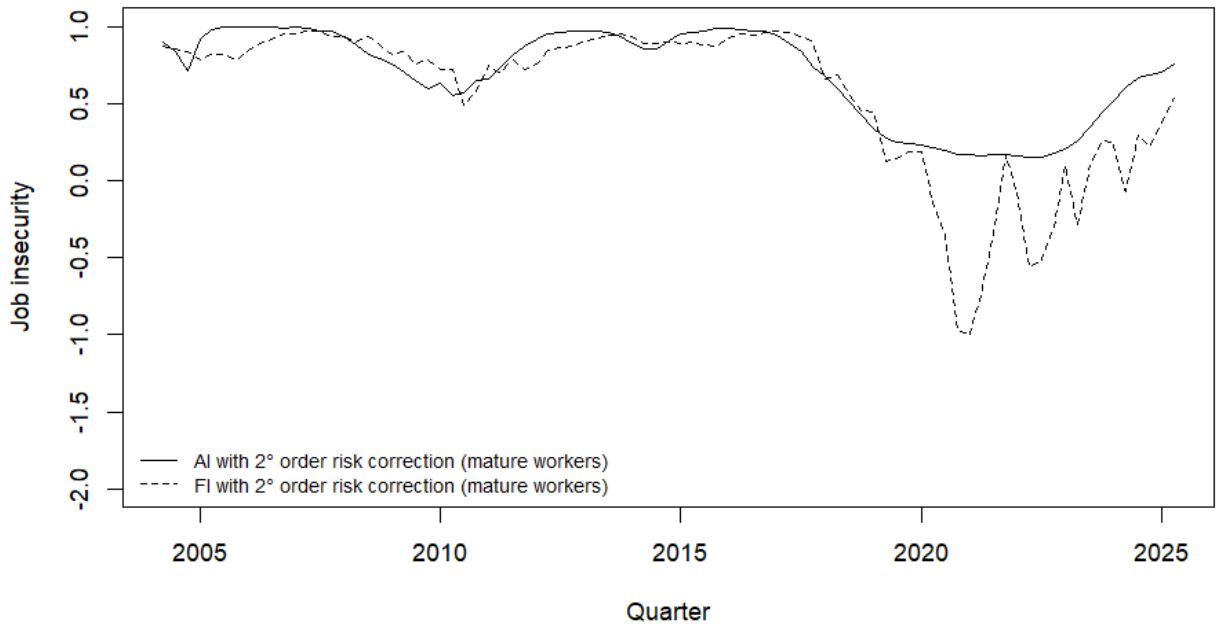

**Table E** Estimated parameters of log-volatility with cohort heterogeneity

| *(1)* Parameter | *(2)* $JI_t^{[0,12],FI\ mature}$ | *(3)* $JI_t^{[0,12],AI\ mature}$ | *(4)* $\tilde{JI}_t^{[0,12],SHARE}$ |
|---|---|---|---|
| $\mu_v$ | -0.7330 {0.5880} | -0.7100 {0.8000} | -6.6910 {1.5960} |
| $\varphi_v$ | 0.9240 {0.0704} | 0.9400 {0.0408} | 0.8730 {0.0730} |
| $\sigma_\eta$ | 0.1960 {0.0980} | 0.3100 {0.0980} | 0.8340 {0.1380} |
| $e^{\frac{\mu_v^2}{2}}$ | 0.7230 {0.2330} | 0.7700 {0.5330} | 0.0830 {0.4540} |
| $\sigma_v^2$ | 0.0480 {0.0406} | 0.1100 {0.0650} | 0.7140 {0.2380} |

Standard deviations in curly brackets

**Figure D** Survey-based index of perceived job insecurity of mature workers

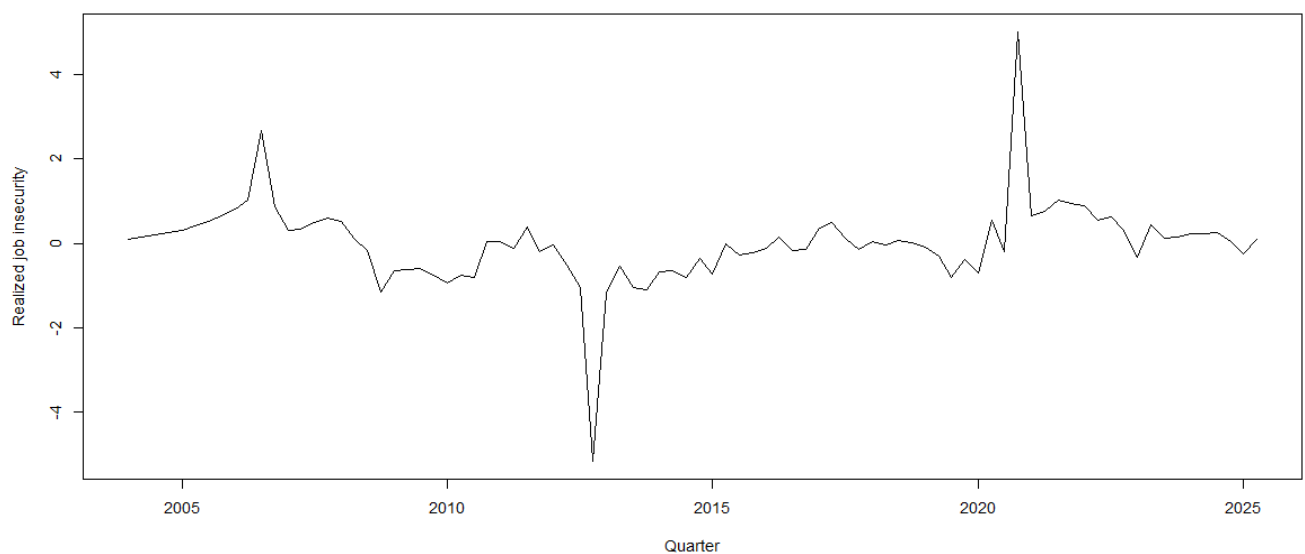

**Table F** Estimation output of equation (62) for mature workers

| (1)<br>Estimated coefficient/test | (2)<br>OLS regression with the "mean-only" job insecurity measures for mature workers | (3)<br>OLS regression with the 2° order risk correction job insecurity measures for mature workers |
|---|---|---|
| $\hat{\alpha}_0$ | 0.0000<br>(0.1861) | 0.0000<br>(0.1186) |
| $\hat{\alpha}_1$ | -0.7350**<br>(0.2733) | -0.73349**<br>(0.2724) |
| $\hat{\alpha}_2$ | 0.4422*<br>(0.2623) | 0.4386*<br>(0.2618) |
| Adjusted $R^2$ | 0.1366 | 0.1376 |
| HAC standard errors | YES | YES |
| Variance Inflation Factor associated with $JI_t^{[0,12],FI}$ | 4.8156 | 4.7822 |
| Variance Inflation Factor associated with $JI_t^{[0,12],AI}$ | 4.8156 | 4.7822 |
| ADF test on $\widetilde{JI}_t^{[0,12],SHARE}$ | -2.5722<br>[0.1085] | -2.8414<br>[0.7134] |
| ADF test on $JI_t^{[0,12],FI\ mature}$ | -1.9750<br>[0.6144] | -1.9919<br>[0.6052] |
| ADF test on $JI_t^{[0,12],AI\ mature}$ | -2.3741<br>[0.3932] | -2.3772<br>[0.3915] |
| ADF test on residuals | -6.7486<br>[0.0000] | -6.7551<br>[0.0000] |
| Ramsey RESET test | 1.6003<br>[0.2082] | 1.3676<br>[0.2610] |
| CUSUM test | -1.0094<br>[0.3158] | -1.0043<br>[0.3182] |
| Martingale difference<br>sequence test | 10.8000<br>[0.4400] | 10.8000<br>[0.4700] |

*, **, and *** indicate, respectively, 0.10, 0.05, and 0.01 significance levels
Standard errors in round brackets
p-values in squared brackets

**Table G** Mean and variance of the observed and simulated measures of job insecurity

| Variable | Mean | Standard deviation |
|---|---|---|
| $JI_t^{[0,12],FI}$ | 0.4093 | 0.6059 |
| $JI_t^{[0,12],AI}$ | 0.5039 | 0.4142 |
| $JI_t^{[0,12],ISTAT}$ | 1.5399 | 13.7809 |
| $JI_t^{[0,12],FI\ mature}$ | 0.56044 | 0.4925 |
| $JI_t^{[0,12],AI\ mature}$ | 0.6910 | 0.2974 |
| $JI_t^{[0,12],SHARE}$ | 0.0027 | 0.0462 |